\documentclass[10pt, journal,twoside]{IEEEtran} 

\ifCLASSOPTIONcompsoc
\usepackage[nocompress]{cite}
\else
\usepackage{cite}
\fi

\usepackage{epsfig}
\usepackage{subfigure}
\usepackage{amsmath}
\usepackage{amssymb}
\usepackage{cases}
\usepackage{hhline}
\usepackage{times}
\usepackage[font=footnotesize,labelfont=bf]{caption}
\usepackage{caption}
\usepackage{algorithm}
\usepackage{algorithmic}
\usepackage[T1]{fontenc}
\usepackage{color}
\usepackage{tabularx}
\usepackage{booktabs}
\usepackage{makecell}
\usepackage{multirow}
\usepackage{flushend}
\usepackage{float}
\usepackage{graphicx}
\usepackage{url}

\begin{document}
\title{Edge-Cloud Collaborative Pothole Detection via Onboard Event Screening and Federated Temporal Segmentation}

\author{Yingjie Wu, Kongyang Chen, Tiancai Liang
\IEEEcompsocitemizethanks{
\IEEEcompsocthanksitem Yingjie Wu, Kongyang Chen, and Tiancai Liang are with School of Artificial Intelligence, Guangzhou University, Guangzhou 510006, China.  
}}

\IEEEtitleabstractindextext{
\begin{abstract}
Road potholes threaten driving safety and increase infrastructure maintenance costs, while large-scale and timely pothole detection remains challenging in urban road networks. Vehicle-mounted vibration sensing offers a low-cost and scalable solution, however, continuous transmission of raw acceleration streams causes high communication overhead. Also, vibration patterns induced by potholes are often confused with those caused by manholes, speed bumps, and other local road structures. To address these challenges, this paper proposes an edge-cloud collaborative pothole detection framework based on onboard vibration event screening and federated temporal segmentation. At the vehicle side, a Gaussian Mixture Model (GMM)-based module adaptively models background vibration and screens candidate abnormal events from continuous acceleration streams. The onboard module acts as a lightweight high-recall filter and uploads only compact candidate event segments with their contextual information. At the server side, pothole detection is formulated as a point-wise temporal segmentation task. A 1D Attention U-Net is developed to distinguish potholes from vibration-similar road events by capturing multi-scale temporal features and preserving event boundary information. Furthermore, the model is trained under a federated learning framework to exploit distributed multi-vehicle data while accommodating non-IID vehicle data distributions. Experiments on multi-vehicle vibration sensing data show that the proposed framework reduces unnecessary data transmission from smooth road segments and improves fine-grained pothole detection under both centralized and federated settings.
\end{abstract}
}

\maketitle
\IEEEdisplaynontitleabstractindextext
\IEEEpeerreviewmaketitle

%-------------------
\section{Introduction}

Road potholes are common road surface defects that may cause vehicle damage, reduce driving comfort, increase accident risks, and impose additional maintenance cost on road operators~\cite{mednis2012embedded}. Timely pothole detection is therefore important for intelligent transportation systems and urban infrastructure management. Conventional inspection methods usually rely on manual patrols or dedicated inspection vehicles equipped with cameras, laser scanners, or other dedicated sensors. Although these methods can provide reliable detection results, they are often labor-intensive, costly, and difficult to deploy continuously over large-scale road networks.

Vehicle-mounted vibration sensing has emerged as a promising alternative for scalable road surface monitoring. By installing lightweight sensing terminals on ordinary vehicles, acceleration signals, GPS locations, vehicle speeds, and timestamps can be collected during daily driving~\cite{mohan2008nericell, eriksson2008pothole}. When a vehicle passes over a pothole or other local road irregularity, the interaction between the tire and the road surface produces characteristic vibration responses. Compared with vision-based inspection, vibration-based sensing is less sensitive to illumination, weather, and camera viewpoints, and can be deployed at relatively low cost across a large number of vehicles.

However, accurate pothole detection from vehicle vibration signals remains challenging. First, vehicle vibration is affected by many factors, including vehicle speed, suspension characteristics, sensor installation position, road material, and driving behavior~\cite{ranyal2022road, winet2016crms}. A fixed acceleration threshold is therefore insufficient to robustly detect abnormal road events under diverse driving conditions. Second, potholes are not the only road structures that cause vibration changes. Manholes, speed bumps, bridge joints, pavement seams, and rough road patches may produce vibration patterns similar to potholes, making lightweight onboard classification unreliable. Third, continuously uploading raw acceleration streams from many vehicles to the cloud would introduce substantial communication and storage overhead, since most driving data correspond to smooth road segments.

To address these issues, this paper proposes an edge-cloud collaborative pothole detection framework based on onboard vibration event screening and server-side federated temporal segmentation. The key idea is to decompose pothole detection into two stages. At the vehicle side, the onboard terminal performs lightweight candidate event screening instead of highly complex pothole classification. Specifically, a Gaussian Mixture Model (GMM) is used to adaptively model background vibration patterns of smooth road segments. Samples that deviate from the learned background distribution are grouped into candidate vibration abnormal events. Only these compact event segments, together with their GPS, speed, timestamp, and vehicle information, are uploaded to the server. This approach reduces unnecessary data transmission while maintaining high recall for potential road events.

At the server side, the uploaded candidate events are further analyzed by a fine-grained temporal segmentation model. Since candidate abnormal events may correspond to potholes, manholes, speed bumps, or other disturbances, the server-side model must learn discriminative temporal patterns rather than simply detect vibration peaks. We formulate pothole detection as a point-wise temporal segmentation problem, where each sampling point in a candidate vibration sequence is assigned a semantic label. A 1D Attention U-Net is developed to perform this task. Its encoder--decoder structure captures multi-scale temporal features, skip connections help preserve event boundary information, and attention modules enhance anomaly-related features while suppressing background noise.

In practical multi-vehicle sensing scenarios, vibration data are naturally distributed across vehicles and exhibit non-independent and identically distributed (non-IID) characteristics. Different vehicles may travel on different routes, encounter different event types, move at different speeds, and have different sensor installation conditions. To exploit such distributed vehicle data without directly centralizing all raw sensing streams, the proposed 1D Attention U-Net is further integrated into a federated learning framework. Each vehicle or vehicle-associated client trains the temporal segmentation model using its local candidate event data, while the server aggregates model updates to obtain a global model. This design improves the scalability of the system and makes the learning process more suitable for multi-vehicle road monitoring.

The main contributions of this paper are summarized as follows:
\begin{itemize}
    \item We propose an edge-cloud collaborative pothole detection framework that combines lightweight onboard event screening with server-side fine-grained temporal recognition. The framework reduces redundant transmission of smooth-road vibration data while preserving potential pothole-related events for further analysis.

    \item We design a GMM-based onboard vibration abnormal event screening method that adaptively models background vibration signals and identifies candidate road events under varying driving conditions. The module is used as a high-recall filter rather than a final pothole classifier.

    \item We develop a federated 1D Attention U-Net for server-side pothole detection. The model formulates pothole recognition as a point-wise temporal segmentation task and supports collaborative learning from distributed non-IID vehicle data.
\end{itemize}

The remainder of this paper is organized as follows. Section~II reviews related studies for vehicular sensing. Section~III presents the edge-cloud collaborative framework and problem formulation. Section~IV introduces the GMM-based onboard event screening method. Section~V describes the federated 1D Attention U-Net for server-side pothole detection. Section~VI presents the experimental setup, and Section~VII reports the experimental results. Finally, Section~VIII concludes this paper and discusses future work.

%-------------------
\section{Related Work}
\subsection{Vehicle Vibration-Based Road Anomaly Detection}

Vehicle vibration-based road anomaly detection has been studied as a low-cost alternative to manual inspection and dedicated road inspection vehicles~\cite{mednis2012embedded}. Early mobile sensing systems used accelerometers and GPS sensors installed on vehicles or smartphones to detect road surface irregularities~\cite{china2021, luo2020road}. For example, Pothole Patrol used vehicle-mounted sensors to identify road surface conditions through opportunistic mobile sensing~\cite{eriksson2008pothole}, while Nericell demonstrated the feasibility of using smartphones to monitor road and traffic conditions~\cite{mohan2008nericell}. Mednis et al. further evaluated real-time pothole detection algorithms using Android smartphones with accelerometers~\cite{mednis2012embedded}. These studies established the basic feasibility of using vibration and location data for scalable road monitoring.

Subsequent studies improved road anomaly detection by using signal processing and machine learning methods~\cite{seraj2014roads, euc2013crms}. Seraj et al. proposed RoADS, which used smartphone inertial sensors and extracted time-domain, frequency-domain, and wavelet-based features for road pavement anomaly detection~\cite{seraj2014roads}. These methods are computationally efficient and suitable for lightweight deployment, but their performance is often sensitive to vehicle speed, sensor placement, road texture, and manually selected thresholds or features. A recent review also shows that road condition monitoring systems have increasingly combined contact sensors, mobile sensing, and artificial intelligence methods, but robustness across heterogeneous real-world conditions remains a key challenge~\cite{ranyal2022road}.

Deep learning has been introduced to reduce the dependence on handcrafted features~\cite{varona2020deep, luo2020road}. Varona et al. investigated deep learning models for automatic road surface monitoring and pothole detection using smartphone sensing data~\cite{varona2020deep}. Recently, edge computing and hybrid models have further improved performance on resource-constrained devices\cite{anand2025assessment, bibi2021edge}. Compared with threshold-based or feature-engineering-based methods, deep models can learn more discriminative representations from vibration signals. 

However, many existing vibration-based methods still treat road event recognition as segment-level classification. This formulation may be insufficient when a candidate sequence contains background vibration, multiple local disturbances, or road events with unclear boundaries. In addition, directly uploading continuous raw vibration streams from a large number of vehicles may introduce unnecessary communication and storage costs. Therefore, this paper adopts a two-stage design: the onboard GMM module performs lightweight high-recall event screening, while the server-side model performs fine-grained temporal segmentation.

\subsection{Temporal Segmentation for Vibration Signals}

Road pothole detection from vibration signals requires not only event classification but also temporal localization. Potholes, manholes, and speed bumps may all produce abrupt acceleration changes, and their differences may lie in waveform duration, impact shape, and post-event oscillation rather than in peak amplitude alone. Therefore, assigning a single label to an entire vibration segment may lose important boundary information. A point-wise temporal segmentation formulation is more suitable for identifying both the category and temporal interval of each road event.

Encoder--decoder networks provide an effective architecture for dense prediction. U-Net was originally proposed for biomedical image segmentation and has become a representative architecture for pixel-wise prediction because of its symmetric encoder--decoder structure and skip connections~\cite{soother2021novel, xie2024elevator}. Although U-Net was first developed for two-dimensional images, its design principle can be naturally extended to one-dimensional temporal signals (e.g., U-TSS\cite{shan2025u}). In vibration-based road event detection, the encoder can extract multi-scale temporal features, while the decoder and skip connections help recover fine-grained boundary information.

Attention mechanisms have also been widely used to improve feature selection in dense prediction networks. Attention U-Net introduced attention gates to suppress irrelevant responses and highlight task-relevant structures~\cite{oktay2018attention}. CBAM further showed that channel and spatial attention can be used as lightweight modules for adaptive feature refinement~\cite{woo2018cbam}. Inspired by these ideas, this paper develops a 1D Attention U-Net for candidate vibration event segmentation. Different from segment-level classifiers, the proposed model outputs point-wise labels and therefore supports event boundary localization and event-level pothole reporting.

\subsection{Federated Learning for Distributed Vehicular Sensing}

Large-scale road monitoring naturally involves data collected from many vehicles. A centralized training strategy can aggregate all vehicle data at the server, but continuous raw-data uploading may cause high communication and storage overhead. In addition, vehicle data are distributed across different routes, vehicle types, driving speeds, and sensor installation conditions. This makes the collected data heterogeneous and non-IID.

Federated learning provides a distributed learning paradigm in which clients train models locally and upload model updates rather than raw data~\cite{mcmahan2017communication, isa2023fl}. FedAvg is a representative federated optimization method that aggregates locally trained models through weighted averaging~\cite{mcmahan2017communication}. Yang et al. summarized major advances and open problems in federated learning, including statistical heterogeneity, communication efficiency, and systems constraints~\cite{tist2019flsurvey}. 
FedProx further studied federated optimization under heterogeneous networks and non-IID data distributions~\cite{li2020fedprox}. Federated learning has been applied to anomaly detection in IoT and industrial scenarios\cite{wang2023federated, shubyn2022federated, xiang2024federated}, with its core advantage being that data does not need to leave local devices for collaborative model training. These studies indicate that federated learning is suitable for distributed edge and mobile sensing scenarios, but non-IID data remains a central challenge.

For vehicle vibration-based pothole detection, federated learning is particularly relevant because different vehicles may observe different road event distributions. Some vehicles may collect more pothole samples, while others may mainly encounter manholes, speed bumps, or normal road segments. Existing federated learning studies mainly focus on general classification or anomaly detection, while federated temporal segmentation for road event recognition has received less attention. This paper integrates a 1D Attention U-Net into a federated learning framework, enabling the server-side model to learn from distributed multi-vehicle candidate events while preserving dense temporal prediction capability.

\subsection{Summary}

Existing vibration-based road monitoring studies demonstrate the feasibility of using accelerometer and GPS data for scalable road anomaly detection. However, lightweight onboard methods are often insufficient for distinguishing potholes from vibration-similar road structures. Deep learning methods improve representation learning, but many existing approaches focus on segment-level classification and do not explicitly model point-wise event boundaries. Federated learning enables collaborative training over distributed vehicle data, but its integration with temporal segmentation for pothole detection is still underexplored. To address these limitations, this paper proposes an edge-cloud collaborative framework that combines GMM-based onboard candidate event screening with federated 1D Attention U-Net-based server-side temporal segmentation.

%-------------------
\section{System Overview and Problem Formulation}

This section presents the overall edge-cloud collaborative framework and formulates the pothole detection task. 

\subsection{edge-cloud Collaborative Architecture}

\begin{figure*}[!t]
    \centering
    \includegraphics[width=0.8\linewidth]{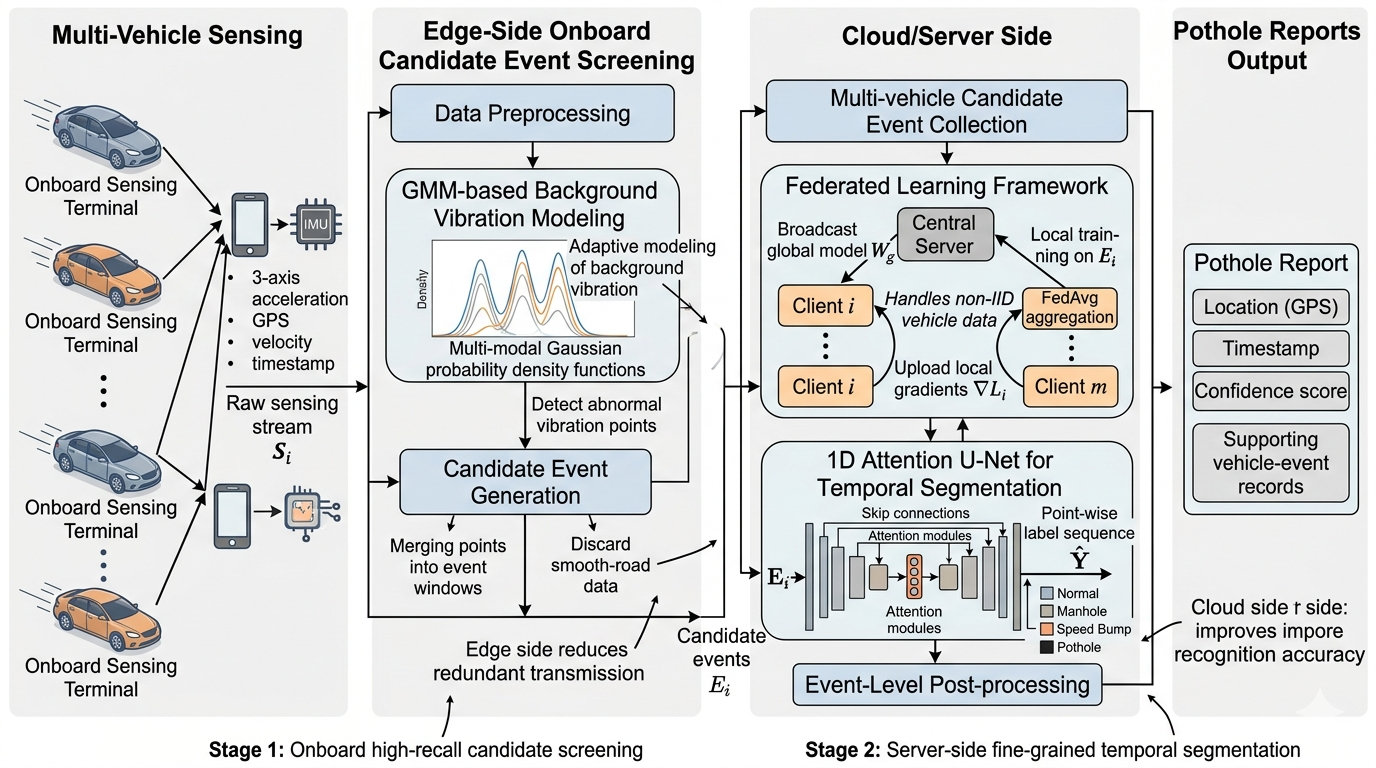}
    \caption{Overview of the proposed edge-cloud collaborative road pothole detection framework..}
    \label{fig:system_architecture}
\end{figure*}

The overall architecture of the proposed framework is shown in Fig.~\ref{fig:system_architecture}. Each vehicle is equipped with an onboard sensing terminal that continuously collects three-axis acceleration, GPS location, vehicle velocity, and timestamp information during normal driving. Instead of uploading the complete raw sensing stream to the cloud, the terminal performs lightweight vibration abnormal event screening locally. Smooth-road segments are discarded, and only candidate vibration event segments are transmitted to the server together with their contextual information.

The edge-side module is designed for communication-efficient candidate event extraction. It uses a GMM-based background vibration model to identify signals that deviate from normal vibration patterns. The output of this module is not a confirmed pothole label, but a set of candidate abnormal events. These events may correspond to potholes, manholes, speed bumps, bridge joints, pavement seams, or other local disturbances.

The cloud-side module is responsible for fine-grained pothole recognition. After receiving candidate events from multiple vehicles, the server organizes them according to vehicle identifiers, timestamps, GPS locations, and event metadata. A 1D Attention U-Net is then used to perform point-wise temporal segmentation over candidate vibration sequences. The segmentation results are further converted into event-level pothole detection outputs through post-processing.

To exploit distributed vehicle data under practical multi-vehicle sensing conditions, the server-side segmentation model is trained within a federated learning framework. Each vehicle or vehicle-associated client trains the model using its local candidate event data and uploads model updates to the server. The server aggregates the updates to obtain a global model. This design allows the system to learn from heterogeneous vehicle data while avoiding direct centralization of all raw sensing streams.

\subsection{Problem Formulation}

Consider a series of $N$ vehicles equipped with onboard sensing terminals. For vehicle $i$, the raw sensing stream is denoted as
\begin{equation}
S_i=\{(\mathbf{a}_i^t,g_i^t,v_i^t,t)\}_{t=1}^{T_i},
\end{equation}
where $\mathbf{a}_i^t=[a_{x,i}^t,a_{y,i}^t,a_{z,i}^t]$ is the three-axis acceleration, $g_i^t$ is the GPS location, $v_i^t$ is the vehicle velocity, and $T_i$ is the length of the sensing stream. The objective is to detect pothole events from the multi-vehicle sensing streams and generate event-level pothole reports with location and time information.

Directly uploading all raw sensing streams $\{S_i\}_{i=1}^{N}$ to the server is inefficient because most driving data correspond to normal road segments. Therefore, the vehicle-side task is formulated as candidate vibration event screening:
\begin{equation}
\mathcal{F}_{\mathrm{edge}}: S_i \rightarrow E_i,
\end{equation}
where
\begin{equation}
E_i=\{e_i^1,e_i^2,\ldots,e_i^{M_i}\}
\end{equation}
is the set of candidate events detected by vehicle $i$, and $M_i$ is the number of candidate events. Each candidate event is represented as
\begin{equation}
e_i^m=(X_i^m,g_i^m,v_i^m,\tau_i^m,id_i),
\end{equation}
where $X_i^m$ is the extracted vibration segment, $g_i^m$ is the associated GPS location, $v_i^m$ is the vehicle velocity, $\tau_i^m$ is the timestamp, and $id_i$ is the vehicle identifier.

The server receives candidate events from multiple vehicles:
\begin{equation}
E=\bigcup_{i=1}^{N}E_i.
\end{equation}
Since different vehicles travel along different routes and experience different driving conditions, the local event data are naturally heterogeneous. Let the local dataset of vehicle $i$ be
\begin{equation}
D_i=\{(X_i^m,Y_i^m)\}_{m=1}^{M_i},
\end{equation}
where $Y_i^m$ is the point-wise ground-truth label sequence of $X_i^m$. In practice, the local data distributions are usually non-IID:
\begin{equation}
P_i(X,Y)\neq P_j(X,Y), \quad i\neq j.
\end{equation}

For a candidate vibration sequence $X\in \mathbb{R}^{C\times L}$, where $C$ is the number of input channels and $L$ is the sequence length, the server-side model predicts a point-wise label sequence:
\begin{equation}
\hat{Y}=\{\hat{y}_1,\hat{y}_2,\ldots,\hat{y}_L\}.
\end{equation}
Each sampling point belongs to one of the following classes:
\begin{equation}
\mathcal{C}=\{\mathrm{normal},\mathrm{manhole},\mathrm{speed\ bump},\mathrm{pothole}\}.
\end{equation}
Thus, the server-side detection task is formulated as dense temporal prediction:
\begin{equation}
\mathcal{F}_{\mathrm{server}}: X \rightarrow \hat{Y}, \quad \hat{y}_l\in \mathcal{C}.
\end{equation}

The point-wise prediction sequence is further converted into event-level pothole results. The final output of the system is a set of pothole reports:
\begin{equation}
\mathcal{P}=\{p^1,p^2,\ldots,p^{N_p}\},
\end{equation}
where each report contains the detected pothole location, timestamp, confidence score, and supporting vehicle-event records.

\subsection{Two-Stage Detection Strategy}

The proposed framework follows a two-stage detection strategy. The first stage on raw sensing streams performs onboard abnormal event screening to generate candidate vibration events.
This stage emphasizes high recall and communication reduction. It removes redundant smooth-road data and preserves potential road-event segments.
The second stage on candidate vibration events performs server-side temporal segmentation to produce event-level pothole reports.
This stage emphasizes fine-grained recognition and event boundary localization. By combining GMM-based onboard screening with federated 1D Attention U-Net-based temporal segmentation, the proposed framework balances edge-side efficiency and server-side recognition accuracy.

%-------------------
\section{Onboard GMM-Based Vibration Event Screening}

The onboard screening module is the first stage of the proposed edge-cloud collaborative framework. Its objective is not to directly determine whether a road event is a pothole, but to identify candidate vibration abnormal events from continuous driving data. Smooth-road vibration segments are discarded locally, while candidate event segments with contextual information are uploaded to the server for fine-grained recognition.

\subsection{Onboard Sensing and Data Preprocessing}

Each vehicle is equipped with an onboard sensing terminal that continuously records three-axis acceleration, GPS location, vehicle velocity, and timestamp information. For a vehicle, the sensing stream is represented as
\begin{equation}
S=\{(t, g^t, v^t, \mathbf{a}^t)\}_{t=1}^{T},
\end{equation}
where $g^t$ denotes the GPS location, $v^t$ denotes the vehicle velocity, and $\mathbf{a}^t=[a_x^t,a_y^t,a_z^t]$ denotes the three-axis acceleration. Since road surface condition directly affect vertical vehicle movement, the vertical acceleration $a_z^t$ is used as the primary signal for onboard background vibration modeling.

Before event screening, the terminal performs lightweight preprocessing to remove invalid records and synchronize sensing information. Records with missing or unreliable GPS information, abnormal timestamps, or incomplete acceleration values are discarded or corrected when the missing interval is short~\cite{tmc2021satprobe, tmc2017Crowdsourced, tosn2019bikegps}. It is to ensure that each uploaded candidate event can be associated with valid acceleration, location, speed, and timestamp information.

\subsection{GMM-Based Background Vibration Modeling}

Vehicle vibration on smooth road segments is not constant. It may vary with road surface, vehicle speed, engine vibration, and sensor noise. Therefore, a single fixed threshold is insufficient to robustly distinguish normal vibration from abnormal road-induced vibration. To adaptively model the background vibration distribution, we use a Gaussian Mixture Model (GMM) on the onboard terminal.

Let $x_t=a_z^t$ denote the vertical acceleration sample at time $t$. The background vibration distribution is modeled as a mixture of $K$ Gaussian components:
\begin{equation}
p(x_t)=\sum_{k=1}^{K}\omega_{k,t}\mathcal{N}(x_t;\mu_{k,t},\sigma_{k,t}^{2}),
\end{equation}
where $\omega_{k,t}$, $\mu_{k,t}$, and $\sigma_{k,t}^{2}$ denote the weight, mean, and variance of the $k$-th Gaussian component, respectively. The component weights satisfy
\begin{equation}
\sum_{k=1}^{K}\omega_{k,t}=1.
\end{equation}

For a newly observed sample $x_t$, the model first checks whether it matches an existing Gaussian component. A component $k$ is regarded as matched if
\begin{equation}
|x_t-\mu_{k,t-1}| \leq M_{\mathrm{match}}\sigma_{k,t-1},
\end{equation}
where $M_{\mathrm{match}}$ is the matching threshold. If multiple components satisfy this condition, the component with the smallest normalized distance is selected:
\begin{equation}
k^{*}=\arg\min_{k}\frac{|x_t-\mu_{k,t-1}|}{\sigma_{k,t-1}}.
\end{equation}

If a matched component exists, its parameters are updated online:
\begin{align}
\mu_{k^{*},t} &=(1-\rho_t)\mu_{k^{*},t-1}+\rho_t x_t,\\
\sigma_{k^{*},t}^{2} &=(1-\rho_t)\sigma_{k^{*},t-1}^{2}
+\rho_t(x_t-\mu_{k^{*},t})^{2},\\
\omega_{k^{*},t} &=(1-\alpha)\omega_{k^{*},t-1}+\alpha,
\end{align}
where $\alpha$ is the learning rate for component weights and $\rho_t$ is the update rate for the matched Gaussian component. For unmatched components, their means and variances remain unchanged, while their weights are decreased:
\begin{equation}
\omega_{k,t}=(1-\alpha)\omega_{k,t-1}, \quad k\neq k^{*}.
\end{equation}
The component weights are then normalized.

If no component matches $x_t$, the sample is considered insufficiently explained by the current background model. The component with the smallest weight is replaced by a new component initialized with the current sample:
\begin{align}
k^{\dagger}&=\arg\min_{k}\omega_{k,t-1},\\
\mu_{k^{\dagger},t}&=x_t,\\
\sigma_{k^{\dagger},t}^{2}&=\sigma_0^{2},\\
\omega_{k^{\dagger},t}&=\omega_0,
\end{align}
where $\sigma_0^{2}$ and $\omega_0$ are the initial variance and weight, respectively. This replacement mechanism enables the GMM to adapt to changing background vibration patterns while preventing occasional abnormal samples from dominating the model.

After model updating, Gaussian components are sorted according to the ratio
\begin{equation}
r_{k,t}=\frac{\omega_{k,t}}{\sigma_{k,t}},
\end{equation}
where components with large weights and small variances are more likely to represent stable background vibration. The first $B$ components are selected as background components such that
\begin{equation}
B=\arg\min_{b}\left(\sum_{k=1}^{b}\omega_{k,t}>T_b\right),
\end{equation}
where $T_b$ is the cumulative background weight threshold. The selected background component set is denoted as $\mathcal{B}_t$.

A sample is identified as an abnormal vibration point if it cannot be explained by any selected background component:
\begin{equation}
I_t=
\begin{cases}
1, & \min\limits_{k\in \mathcal{B}_t}
\frac{|x_t-\mu_{k,t}|}{\sigma_{k,t}}>M_{\mathrm{event}},\\
0, & \mathrm{otherwise},
\end{cases}
\end{equation}
where $M_{\mathrm{event}}$ is the event detection threshold. Here, $I_t=1$ indicates that the current sample is a potential abnormal vibration point, while $I_t=0$ indicates that it belongs to the background vibration.

\subsection{Candidate Event Generation}

The GMM produces a point-level abnormality indicator sequence $\{I_t\}$. Since a real road event usually lasts for a short temporal interval rather than a single sample, adjacent abnormal points are grouped into candidate vibration events. Consecutive abnormal intervals are first identified from $\{I_t\}$ and neighboring intervals are merged if their temporal gap is smaller than a predefined threshold $\Delta_{\mathrm{merge}}$. Very short intervals are removed to suppress isolated sensor spikes or transient noise.

For each retained abnormal interval $[t_s^m,t_e^m]$, the terminal extracts an acceleration segment with left and right temporal context:
\begin{equation}
X^m=\{\mathbf{a}^{t_s^m-w_l},\ldots,\mathbf{a}^{t_e^m+w_r}\},
\end{equation}
where $w_l$ and $w_r$ denote the numbers of samples preserved before and after the detected interval. The added context helps preserve the complete waveform transition around the abnormal vibration, which is important for distinguishing potholes from vibration-similar events at the server side.

The corresponding GPS location, vehicle speed, timestamp, and vehicle identifier are attached to the extracted segment. The candidate event packet is represented as
\begin{equation}
e^m=(X^m,g^m,v^m,\tau^m,id),
\end{equation}
where $g^m$ and $v^m$ are obtained from synchronized valid records within the event window, $\tau^m$ is the event timestamp, and $id$ is the vehicle identifier. Candidate events without reliable location information or with incomplete acceleration segments are discarded.

The onboard module therefore converts continuous raw sensing streams into compact candidate event packets:
\begin{equation}
S \rightarrow \mathcal{E}=\{e^1,e^2,\ldots,e^M\}.
\end{equation}
Only $\mathcal{E}$ is uploaded to the server. In this way, the system avoids transmitting large volumes of smooth-road vibration data while retaining potential pothole-related events for server-side temporal segmentation.

\section{Federated 1D Attention U-Net for Pothole Detection}

After onboard event screening, the server receives candidate vibration events from multiple vehicles. These candidate events contain potholes as well as vibration-similar road structures, such as manholes and speed bumps. Therefore, the server-side module aims to perform fine-grained temporal segmentation and identify pothole intervals from candidate vibration sequences. To this end, we develop a 1D Attention U-Net and integrate it into a federated learning framework.

\subsection{Temporal Segmentation Formulation}

For each candidate event, the input vibration sequence is denoted as
\begin{equation}
X \in \mathbb{R}^{C \times L},
\end{equation}
where $C$ is the number of input channels and $L$ is the sequence length. When only vertical acceleration is used, $C=1$; when three-axis acceleration is used, $C=3$. All candidate sequences are resized or padded to a fixed length and normalized before being fed into the model.

Different from segment-level classification, this work formulates pothole detection as a point-wise temporal segmentation task. For an input sequence $X$, the ground-truth label sequence is
\begin{equation}
Y=\{y_1,y_2,\ldots,y_L\},
\end{equation}
where each $y_l$ belongs to the class set
\begin{equation}
\mathcal{C}=\{0,1,2,3\}.
\end{equation}
The four labels correspond to normal background, manhole, speed bump, and pothole, respectively. The server-side model learns a mapping
\begin{equation}
f_{\theta}: \mathbb{R}^{C\times L}\rightarrow \mathbb{R}^{|\mathcal{C}|\times L},
\end{equation}
where $\theta$ denotes model parameters. The predicted class probability matrix is
\begin{equation}
P=f_{\theta}(X),
\end{equation}
and the point-wise prediction is obtained by
\begin{equation}
\hat{y}_l=\arg\max_{c\in \mathcal{C}}P_{c,l}.
\end{equation}
This formulation enables the model to distinguish event categories and locate event boundaries within the same candidate sequence.

\subsection{1D Attention U-Net Architecture}

\begin{figure*}
    \centering
    \includegraphics[width=0.8\linewidth]{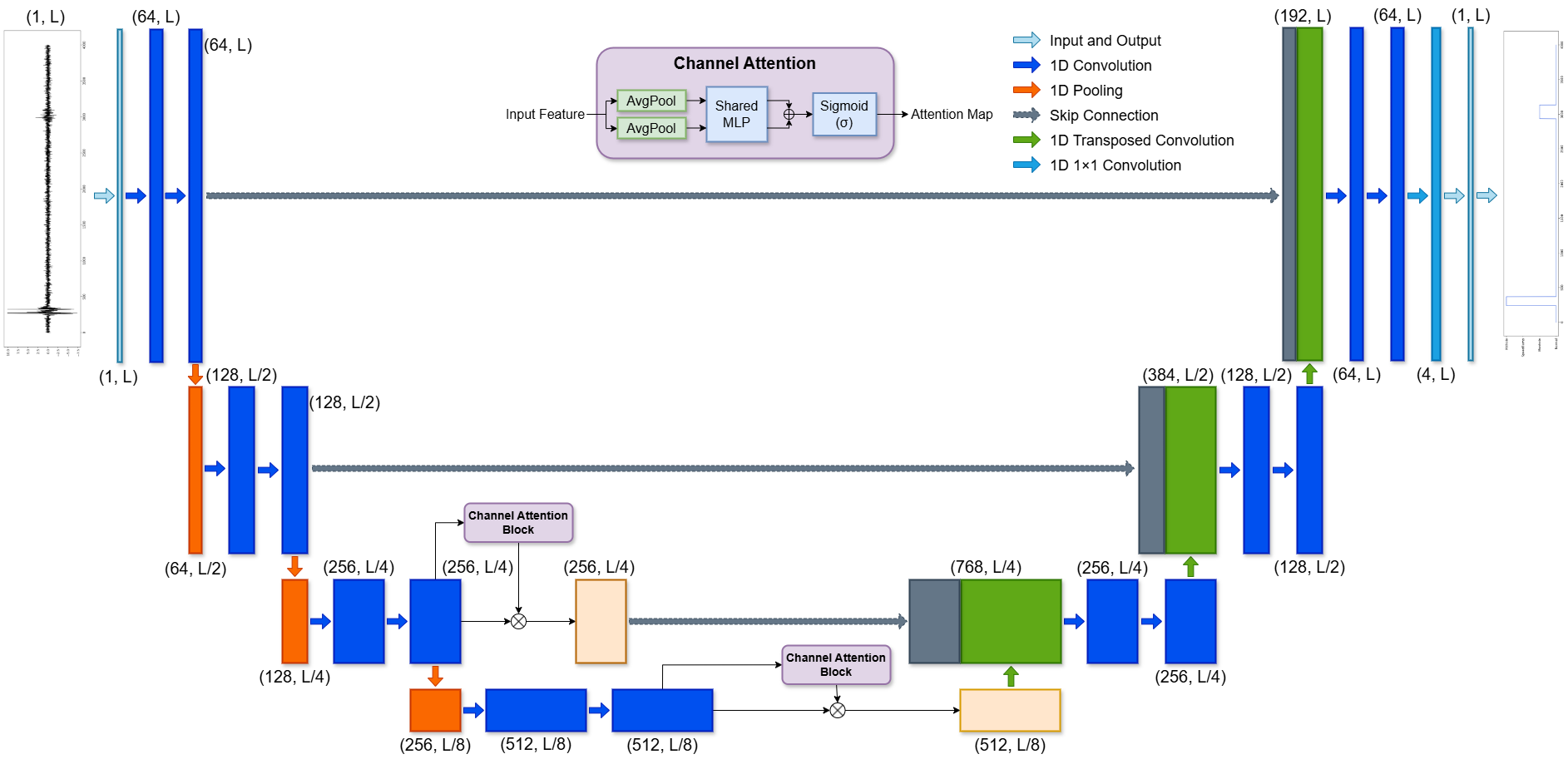}
    \caption{Overall architecture of the proposed 1D Attention U-Net.}
    \label{fig:arch}
\end{figure*}

The proposed 1D Attention U-Net follows an encoder--decoder architecture with skip connections and attention modules, as shown in Fig.~\ref{fig:arch}. The encoder extracts multi-scale temporal features from the input vibration sequence, while the decoder restores temporal resolution and generates point-wise predictions. Skip connections transfer high-resolution features from the encoder to the decoder, which helps preserve local boundary information.

Given the input sequence $X$, the encoder consists of several one-dimensional convolutional blocks. Each block contains two 1D convolutional layers followed by normalization and nonlinear activation:
\begin{equation}
F^{s}=\mathrm{ConvBlock}(F^{s-1}), \quad F^{0}=X,
\end{equation}
where $F^{s}$ denotes the feature map at the $s$-th encoder stage. Downsampling is applied after each encoder block to enlarge the receptive field and capture higher-level temporal patterns. For vibration-based road event recognition, this multi-scale representation is important because potholes, manholes, and speed bumps may differ not only in peak amplitude but also in waveform duration, impact shape, and post-event oscillation.

To enhance discriminative event-related features, attention modules are inserted into the network. For an intermediate feature map $F\in \mathbb{R}^{C_s\times L_s}$, temporal global average pooling and max pooling are first used to generate compact channel descriptors:
\begin{equation}
z_{\mathrm{avg}}=\mathrm{AvgPool}(F), \quad
z_{\mathrm{max}}=\mathrm{MaxPool}(F).
\end{equation}
The descriptors are passed through a shared bottleneck transformation and combined to obtain channel attention weights:
\begin{equation}
w=\sigma\left(\mathrm{MLP}(z_{\mathrm{avg}})+\mathrm{MLP}(z_{\mathrm{max}})\right),
\end{equation}
where $\sigma(\cdot)$ is the sigmoid function. The feature map is recalibrated as
\begin{equation}
\widetilde{F}=w\otimes F,
\end{equation}
where $\otimes$ denotes channel-wise multiplication. The attention mechanism encourages the model to emphasize channels related to abnormal vibration patterns and suppress irrelevant background responses.

The decoder gradually upsamples deep features to recover the original temporal resolution. At each decoder stage, the upsampled feature map is associated with the corresponding encoder feature map:
\begin{equation}
H^{s}=\mathrm{Up}(D^{s+1})\oplus F^{s},
\end{equation}
where $D^{s+1}$ is the decoder feature map from the deeper stage, $\mathrm{Up}(\cdot)$ denotes upsampling, and $\oplus$ denotes channel-wise concatenation. The concatenated feature is further refined by convolutional blocks:
\begin{equation}
D^{s}=\mathrm{ConvBlock}(H^{s}).
\end{equation}

Finally, a point-wise $1\times 1$ convolution maps the final decoder feature map to the class space:
\begin{equation}
Z=\mathrm{Conv}_{1\times 1}(D^{1}),
\end{equation}
and the softmax function produces the class probability at each sampling point:
\begin{equation}
P_{c,l}=\frac{\exp(Z_{c,l})}{\sum_{c'=1}^{|\mathcal{C}|}\exp(Z_{c',l})}.
\end{equation}
The resulting label sequence is used for both point-wise evaluation and event-level pothole detection.

\subsection{Loss Function for Imbalanced Temporal Segmentation}

The temporal segmentation task is highly imbalanced because most sampling points correspond to normal background vibration, while pothole, manhole, and speed bump regions occupy relatively short intervals. To reduce the dominance of background samples and improve event-level segmentation quality, we adopt a combined Focal--Tversky loss.

The Focal loss focuses training on hard and minority-class samples:
\begin{equation}
\mathcal{L}_{\mathrm{Focal}}
=-\frac{1}{L}\sum_{l=1}^{L}
\alpha_{y_l}(1-P_{y_l,l})^{\gamma}
\log(P_{y_l,l}+\epsilon),
\end{equation}
where $\alpha_{y_l}$ is the class weight, $\gamma$ is the focusing parameter, and $\epsilon$ is a small constant for numerical stability.

The Tversky loss is used to improve the overlap between predicted and ground-truth event regions. For class $c$, the soft true positives, false positives, and false negatives are defined as
\begin{align}
TP_c &= \sum_{l=1}^{L} P_{c,l}Y_{c,l},\\
FP_c &= \sum_{l=1}^{L} P_{c,l}(1-Y_{c,l}),\\
FN_c &= \sum_{l=1}^{L} (1-P_{c,l})Y_{c,l},
\end{align}
where $Y_{c,l}$ is the one-hot ground-truth label. The Tversky index is
\begin{equation}
TI_c=\frac{TP_c+\epsilon}{TP_c+\lambda FP_c+\beta FN_c+\epsilon},
\end{equation}
and the Tversky loss is
\begin{equation}
\mathcal{L}_{\mathrm{Tversky}}
=1-\frac{1}{|\mathcal{C}|}\sum_{c=1}^{|\mathcal{C}|}TI_c.
\end{equation}
The final training loss is defined as
\begin{equation}
\mathcal{L}
=\mathcal{L}_{\mathrm{Focal}}
+\eta \mathcal{L}_{\mathrm{Tversky}},
\end{equation}
where $\eta$ controls the contribution of the Tversky term. This loss helps the model focus on minority road-event regions while preserving event-level segmentation consistency.

\subsection{Federated Optimization over Multi-Vehicle Data}

In multi-vehicle road monitoring, candidate vibration events are naturally distributed across vehicles. Let $D_i$ denote the local candidate event dataset of vehicle $i$, where $i=1,\ldots,N$. Due to different driving routes, speeds, vehicle types, and sensor installation conditions, the local data distributions are usually non-IID. Directly training on centralized raw data may increase communication and storage overhead. Therefore, we train the 1D Attention U-Net under a federated learning framework.

At communication round $r$, the server broadcasts the current global model parameters $\theta^{r}$ to a subset of participating clients $\mathcal{S}^{r}$. Each selected client initializes its local model with $\theta^{r}$ and performs local optimization on its dataset $D_i$:
\begin{equation}
\theta_i^{r+1}
=\theta^{r}-\xi \nabla_{\theta}\mathcal{L}_i(\theta^{r};D_i),
\end{equation}
where $\xi$ is the local learning rate and $\mathcal{L}_i$ is the segmentation loss on client $i$. In practice, each client may perform multiple local epochs before uploading its model update.

After receiving local model parameters from selected clients, the server aggregates them using data-size-weighted averaging:
\begin{equation}
\theta^{r+1}
=\sum_{i\in \mathcal{S}^{r}}
\frac{|D_i|}{\sum_{j\in \mathcal{S}^{r}}|D_j|}
\theta_i^{r+1}.
\end{equation}
The updated global model is then used in the next communication round. Through this iterative process, the model learns from distributed vehicle data without requiring all raw sensing streams to be uploaded to the server.

The federated design is consistent with the two-stage architecture of the proposed system. The onboard GMM module reduces raw-data transmission by extracting candidate events, while federated optimization allows the server-side temporal segmentation model to benefit from multi-vehicle data under heterogeneous distributions.

\subsection{Event-Level Post-Processing}

The 1D Attention U-Net outputs a point-wise label sequence for each candidate vibration segment. However, point-wise predictions may contain isolated noisy labels or very short false segments. Therefore, a lightweight post-processing step is applied to obtain event-level pothole results.

First, isolated labels shorter than a predefined duration threshold are removed or merged into neighboring background regions. Second, consecutive sampling points predicted as the same abnormal class are grouped into event intervals. Third, an interval is reported as a pothole event if its dominant label is pothole and its confidence score exceeds a threshold. The event confidence is computed from the average pothole probability within the predicted interval:
\begin{equation}
s_{\mathrm{pot}}=
\frac{1}{|\Omega|}
\sum_{l\in \Omega}P_{\mathrm{pot},l},
\end{equation}
where $\Omega$ denotes the predicted pothole interval. The detected pothole interval is then associated with the GPS location, timestamp, and vehicle identifier of the corresponding candidate event to generate the final road maintenance report.

This post-processing step converts dense temporal predictions into compact event-level pothole reports, making the model output suitable for practical road monitoring and maintenance applications.

%-------------------
\section{Experimental Setup}

This section describes the experimental setup used to evaluate the proposed edge-cloud collaborative pothole detection framework. 

\subsection{Dataset and Prototype Deployment}

We deploy a prototype vehicle sensing system to collect road vibration data in real driving environments. Each participating vehicle is equipped with an onboard terminal that integrates a three-axis accelerometer, a GPS module, a lightweight processing unit, and a wireless communication module. The terminal continuously records synchronized acceleration, location, velocity, and timestamp information during normal driving. The vehicles travel along different urban roads and encounter different road events, including potholes, manholes, speed bumps, and normal road segments.

The collected dataset is constructed from candidate vibration abnormal events generated by the onboard GMM-based screening module. For each candidate event, the corresponding acceleration segment is extracted and associated with GPS location, vehicle velocity, timestamp, and vehicle identifier. Ground-truth labels are obtained by manual inspection and auxiliary road-surface records. Each candidate sequence is annotated in a point-wise manner, where every sampling point is assigned one of four labels: normal background, manhole, speed bump, or pothole.

For model training, each candidate event is converted into a fixed-length vibration sequence. Short sequences are padded or interpolated, while long sequences are cropped around the abnormal interval. Each input sequence is denoted as
\begin{equation}
X\in \mathbb{R}^{C\times L},
\end{equation}
where $C$ is the number of acceleration channels and $L$ is the unified sequence length. The vertical acceleration is used as the primary input channel, and three-axis acceleration can be used for multi-channel input. Each sequence is normalized before being fed into the model to reduce vehicle-dependent amplitude differences.

The main dataset information is summarized in Table~\ref{tab:dataset_summary}.

\begin{table}[t]
\centering
\caption{Summary of the collected multi-vehicle vibration dataset.}
\label{tab:dataset_summary}
\scriptsize
\begin{tabular}{ll}
\hline
Item & Description \\
\hline
Number of deployed vehicles & 100 \\
Sensor type & Three-axis accelerometer and GPS \\
Input signal & Vehicle vibration acceleration \\
Contextual information & GPS, velocity, timestamp, vehicle ID \\
Road event classes & Normal, manhole, speed bump, pothole \\
Annotation type & Point-wise temporal labels \\
Training settings & Centralized training and federated learning \\
\hline
\end{tabular}
\end{table}

\subsection{Baseline Methods}

To evaluate the proposed server-side temporal segmentation model, we compare it with both traditional and deep learning baselines. The traditional baselines include feature-based classifiers trained on statistical vibration features. The deep learning baselines include representative one-dimensional neural models for time-series classification or segmentation.

The compared methods are as follows:
\begin{itemize}
    \item \textbf{Transformer}: The Transformer baseline uses self-attention to model global temporal dependencies in the vibration sequence. 
    \item \textbf{CNN-Transformer}: This baseline combines convolutional layers for local feature extraction with a Transformer encoder for global dependency modeling. 
    \item \textbf{CNN-LSTM}: This baseline first uses convolutional layers to extract local temporal features and then applies LSTM layers to capture sequential dependencies. 
    \item \textbf{Our proposed model (1D Attention U-Net)}: The proposed 1D Attention U-Net trained under centralized or federated settings.
\end{itemize}

For fair comparison, all deep learning models use the same input sequences and training/testing split. For segment-level baselines, point-wise labels are converted into segment labels according to the dominant event class in each candidate sequence. For temporal segmentation models, the original point-wise labels are used for training and evaluation.

\subsection{Evaluation Metrics}
For point-wise temporal segmentation, we report accuracy, precision, recall, F1-score, and weighted F1-score. These metrics are computed over all sampling points in the test set. Since normal background points dominate the sequences, class-wise and weighted metrics are used to better reflect the detection performance on minority road-event classes.

For event-wise evaluation, continuous non-normal segments are extracted from the post-processed prediction sequence. A predicted event is regarded as correctly matched if its temporal Intersection over Union (IoU) with a ground-truth event exceeds a threshold $\delta_{\mathrm{IoU}}$ and its predicted class is correct. Event-wise precision, recall, and F1-score are computed as
\begin{align}
\mathrm{Precision}_{\mathrm{event}} &=
\frac{N_{\mathrm{matched}}}{N_{\mathrm{pred}}},\\
\mathrm{Recall}_{\mathrm{event}} &=
\frac{N_{\mathrm{matched}}}{N_{\mathrm{gt}}},\\
\mathrm{F1}_{\mathrm{event}} &=
\frac{2\cdot \mathrm{Precision}_{\mathrm{event}}\cdot \mathrm{Recall}_{\mathrm{event}}}
{\mathrm{Precision}_{\mathrm{event}}+\mathrm{Recall}_{\mathrm{event}}},
\end{align}
where $N_{\mathrm{matched}}$, $N_{\mathrm{pred}}$, and $N_{\mathrm{gt}}$ denote the numbers of matched, predicted, and ground-truth events, respectively. We report event-wise results under different IoU thresholds to evaluate localization quality.

\subsection{Implementation Details}

The proposed framework is implemented using a server--client training pipeline. The onboard module performs data cleaning, GMM-based background modeling, and candidate event extraction. The server module performs dataset organization, model training, federated aggregation, inference, and post-processing.

For the GMM-based screening module, the number of Gaussian components is set to $K=4$. The event detection threshold is selected to maintain high recall, because false candidate events can be further filtered by the server-side model. Detected abnormal points are merged into candidate intervals, and short isolated intervals are removed before event packets are generated.

For the 1D Attention U-Net, the model is trained using the Focal--Tversky loss. Adam is used as the optimizer. The initial learning rate is set to 0.001, and cosine annealing is adopted for learning-rate scheduling. The batch size is set to 32. For centralized training, the maximum number of epochs is set to 100, and early stopping is applied according to validation performance.

For federated learning, the server initializes the global model and broadcasts it to selected clients in each communication round. Each selected client trains the model on its local dataset and uploads updated parameters to the server. The server aggregates local models using sample-size-weighted FedAvg. The number of communication rounds is set to 20, and each selected client performs 2 local epochs per round.

The main implementation parameters are summarized in Table~\ref{tab:implementation_details}.

\begin{table}[t]
\centering
\caption{Implementation details of model training.}
\label{tab:implementation_details}
\scriptsize
\begin{tabular}{ll}
\hline
Parameter & Value \\
\hline
Input sequence length & $L$ \\
Number of classes & 4 \\
Batch size & 32 \\
Initial learning rate & 0.001 \\
Optimizer & Adam \\
Centralized training epochs & 100 \\
Federated communication rounds & 20 \\
Federated aggregation & FedAvg \\
GMM components & 4 \\
Post-processing & Mode \& length filtering \\
\hline
\end{tabular}
\end{table}
%-------------------
\section{Experimental Results}

This section reports the experimental results of the proposed edge-cloud collaborative pothole detection framework. 

\subsection{GMM-Based Candidate Event Screening}

The onboard GMM-based module is evaluated on continuous acceleration streams collected by vehicle-mounted terminals. Since this module is designed as a high-recall candidate event screener rather than a final pothole classifier, its main objective is to retain potential road events while reducing the transmission of smooth-road vibration data.

\begin{figure}[!t]
	\centering 
	\subfigure[Z-axis acceleration.]{
	\includegraphics[width=0.4\linewidth]{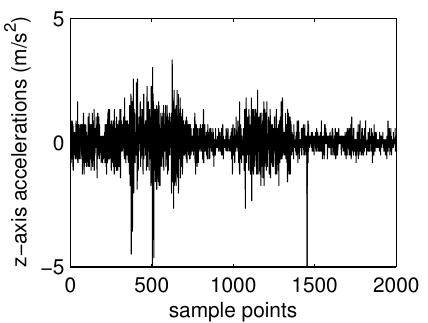}
	\label{subfig:z}
	}
	%\quad%\hspace{10mm}
	\subfigure[Z-peak with Z\_th=1.]{
	\includegraphics[width=0.4\linewidth]{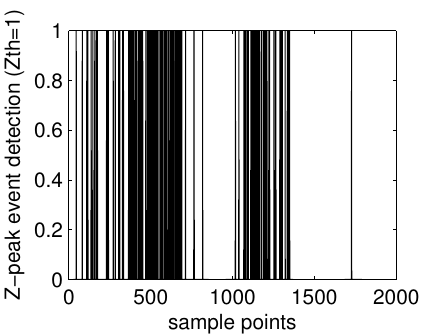}
	\label{subfig:zpeak1}
	}
	%\quad%\hspace{10mm}
	\subfigure[Z-peak with Z\_th=2.2.]{
	\includegraphics[width=0.4\linewidth]{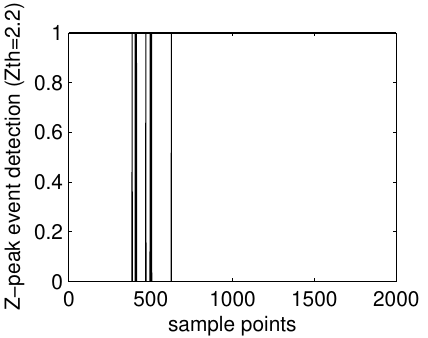}
	\label{subfig:zpeak22}
	}
	%\quad%\hspace{10mm}
	\subfigure[GMM.]{
	\includegraphics[width=0.4\linewidth]{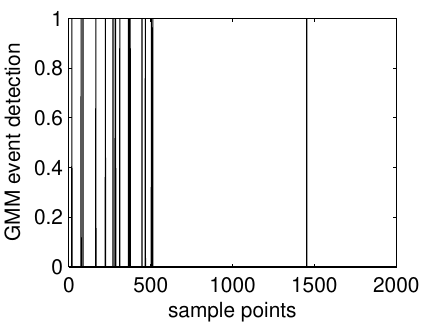}
	\label{subfig:gmm}
	}	
	\caption{Example of GMM-based onboard candidate event screening.}
	\label{fig:gmm_screening}
\end{figure}

Fig.~\ref{fig:gmm_screening} illustrates an example of GMM-based candidate event screening. The vertical acceleration signal is continuously monitored, and the GMM detects vibration samples that deviate from the learned background distribution. Adjacent abnormal samples are then grouped into candidate event intervals. These intervals may correspond to potholes, manholes, speed bumps, or other local road disturbances, and are further analyzed by the server-side temporal segmentation model.

Compared with fixed-threshold triggering methods, the GMM-based module is more adaptive to changes in vehicle speed, road texture, suspension response, and sensor installation conditions. A fixed threshold may either miss weak road events or generate excessive false alarms under different driving conditions. In contrast, the GMM updates the background vibration model online and detects samples that are unlikely to be explained by the selected background components.

\subsection{Centralized Training Results}

We first evaluate the server-side temporal segmentation models under centralized training, where all training samples are available to train a single model. Although the proposed framework is designed for federated learning, centralized training provides a useful upper-bound reference. The proposed 1D Attention U-Net is compared with Transformer, CNN-Transformer, and CNN-LSTM baselines.

Table~\ref{tab:centralized_core_metrics} reports the core point-wise and event-wise metrics. The proposed U-Net achieves the best performance across all major metrics, with an overall point-wise accuracy of 0.9964, a point-wise macro-average F1-score of 0.9656, and an event-level macro-average F1-score of 0.9426. These results show that the proposed model can accurately classify sampling points and identify complete road events.

\begin{table}[t]
\centering
\caption{Comparison of core point-wise and event-wise metrics under centralized training.}
\label{tab:centralized_core_metrics}
\scriptsize
\begin{tabular}{lccccc}
\hline
Model & Acc. & \multicolumn{2}{c}{Point-wise F1} & \multicolumn{2}{c}{Event-level F1} \\
\cline{3-6}
 &  & Macro & Weighted & Macro & Weighted \\
\hline
Transformer & 0.9777 & 0.7912 & 0.9787 & 0.2584 & 0.2591 \\
CNN-Transformer & 0.9923 & 0.9210 & 0.9924 & 0.5247 & 0.5225 \\
CNN-LSTM & 0.9931 & 0.9312 & 0.9932 & 0.8247 & 0.8247 \\
1D Attention U-Net & \textbf{0.9964} & \textbf{0.9656} & \textbf{0.9965} & \textbf{0.9426} & \textbf{0.9424} \\
\hline
\end{tabular}
\end{table}

The Transformer baseline achieves relatively high point-wise accuracy but much lower event-level F1-score. This indicates that correctly classifying dominant background points does not necessarily lead to reliable event detection. CNN-Transformer improves point-wise performance by introducing local convolutional features, but its event-level performance remains limited. CNN-LSTM performs better in event-wise metrics, showing the usefulness of temporal dependency modeling. However, it is still inferior to the proposed U-Net because recurrent modeling does not explicitly recover high-resolution event boundaries.

Table~\ref{tab:centralized_iou} reports event-level localization performance under centralized training. The proposed U-Net achieves the highest mean IoU of 0.9318 and maintains the best F1-score under all IoU thresholds. At the strict threshold of IoU = 0.7, it still obtains an F1-score of 0.9061, indicating that the predicted event intervals are well aligned with ground-truth annotations.

\begin{table}[t]
\centering
\caption{Comparison of event-level localization accuracy under centralized training.}
\label{tab:centralized_iou}
\scriptsize
\begin{tabular}{lccccc}
\hline
Model & Mean IoU & F1@0.1 & F1@0.3 & F1@0.5 & F1@0.7 \\
\hline
Transformer & 0.7877 & 0.2766 & 0.2542 & 0.2374 & 0.1910 \\
CNN-Transformer & 0.8714 & 0.5228 & 0.5143 & 0.4924 & 0.4536 \\
CNN-LSTM & 0.8774 & 0.8314 & 0.8193 & 0.8053 & 0.7570 \\
1D Attention U-Net & \textbf{0.9318} & \textbf{0.9459} & \textbf{0.9407} & \textbf{0.9370} & \textbf{0.9061} \\
\hline
\end{tabular}
\end{table}

These results demonstrate that event-wise evaluation is more informative than point-wise accuracy for pothole detection. For road maintenance applications, the model must generate complete and correctly localized pothole reports rather than only classify isolated sampling points.

%-------------------
\subsection{Federated Learning Results}

We further evaluate the proposed model under federated learning. In this setting, vehicle clients train local models using their own candidate event data, and the server aggregates model parameters using FedAvg. This setting reflects practical multi-vehicle sensing scenarios, where data are distributed and non-IID across vehicles.

Table~\ref{tab:federated_core_metrics} reports the core metrics under federated learning. The proposed U-Net again achieves the best overall performance, with a point-wise accuracy of 0.9969, a point-wise macro-average F1-score of 0.9669, and an event-level macro-average F1-score of 0.9244. Although the training data are distributed across heterogeneous clients, the model still maintains strong segmentation and event recognition capability.

\begin{table}[t]
\centering
\caption{Comparison of core point-wise and event-wise metrics under federated learning.}
\label{tab:federated_core_metrics}
\scriptsize
\begin{tabular}{lccccc}
\hline
Model & Acc. & \multicolumn{2}{c}{Point-wise F1} & \multicolumn{2}{c}{Event-level F1} \\
\cline{3-6}
 &  & Macro & Weighted & Macro & Weighted \\
\hline
Transformer & 0.9691 & 0.7098 & 0.9697 & 0.2167 & 0.2142 \\
CNN-Transformer & 0.9920 & 0.9096 & 0.9922 & 0.5195 & 0.5289 \\
CNN-LSTM & 0.9923 & 0.9175 & 0.9924 & 0.8392 & 0.8425 \\
1D Attention U-Net & \textbf{0.9969} & \textbf{0.9669} & \textbf{0.9969} & \textbf{0.9244} & \textbf{0.9256} \\
\hline
\end{tabular}
\end{table}

The performance gap between U-Net and the baselines remains clear under federated learning. Transformer-based models are weak in event-level detection, especially under non-IID client distributions. CNN-LSTM achieves better event-wise results than Transformer and CNN-Transformer, but it still lacks the dense boundary recovery capability of the encoder--decoder segmentation architecture.

Table~\ref{tab:federated_iou} reports event-level localization accuracy under federated learning. The proposed U-Net achieves the highest mean IoU of 0.9418 and the best F1-scores under all IoU thresholds. At IoU = 0.7, it obtains an F1-score of 0.9025, showing that the federated model can still produce accurately localized event intervals.

\begin{table}[t]
\centering
\caption{Comparison of event-level localization accuracy under federated learning.}
\label{tab:federated_iou}
\scriptsize
\begin{tabular}{cccccc}
\hline
Model & Mean IoU & F1@0.1 & F1@0.3 & F1@0.5 & F1@0.7 \\
\hline
Transformer & 0.7333 & 0.2382 & 0.2094 & 0.1878 & 0.1410 \\
CNN-Transformer & 0.8737 & 0.5296 & 0.5156 & 0.4972 & 0.4682 \\
CNN-LSTM & 0.8691 & 0.8398 & 0.8301 & 0.8135 & 0.7707 \\
1D Attention U-Net & \textbf{0.9418} & \textbf{0.9261} & \textbf{0.9188} & \textbf{0.9099} & \textbf{0.9025} \\
\hline
\end{tabular}
\end{table}

\begin{figure*}[!t]
	\centering 
	\subfigure[U-Net + Point-wise.]{
	\includegraphics[width=0.22\linewidth]{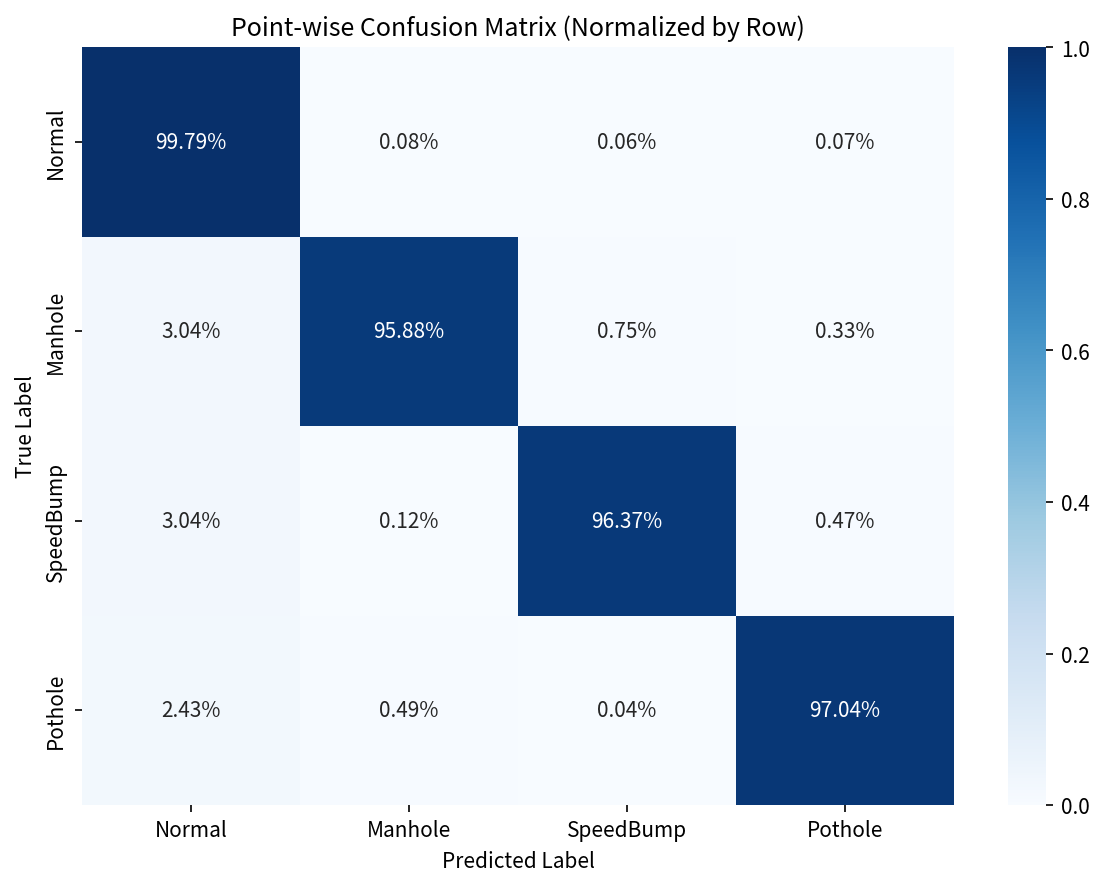}
	\label{subfig:fig20a}
	}
	%\quad%\hspace{10mm}
	\subfigure[CNN-LSTM + Point-wise.]{
	\includegraphics[width=0.22\linewidth]{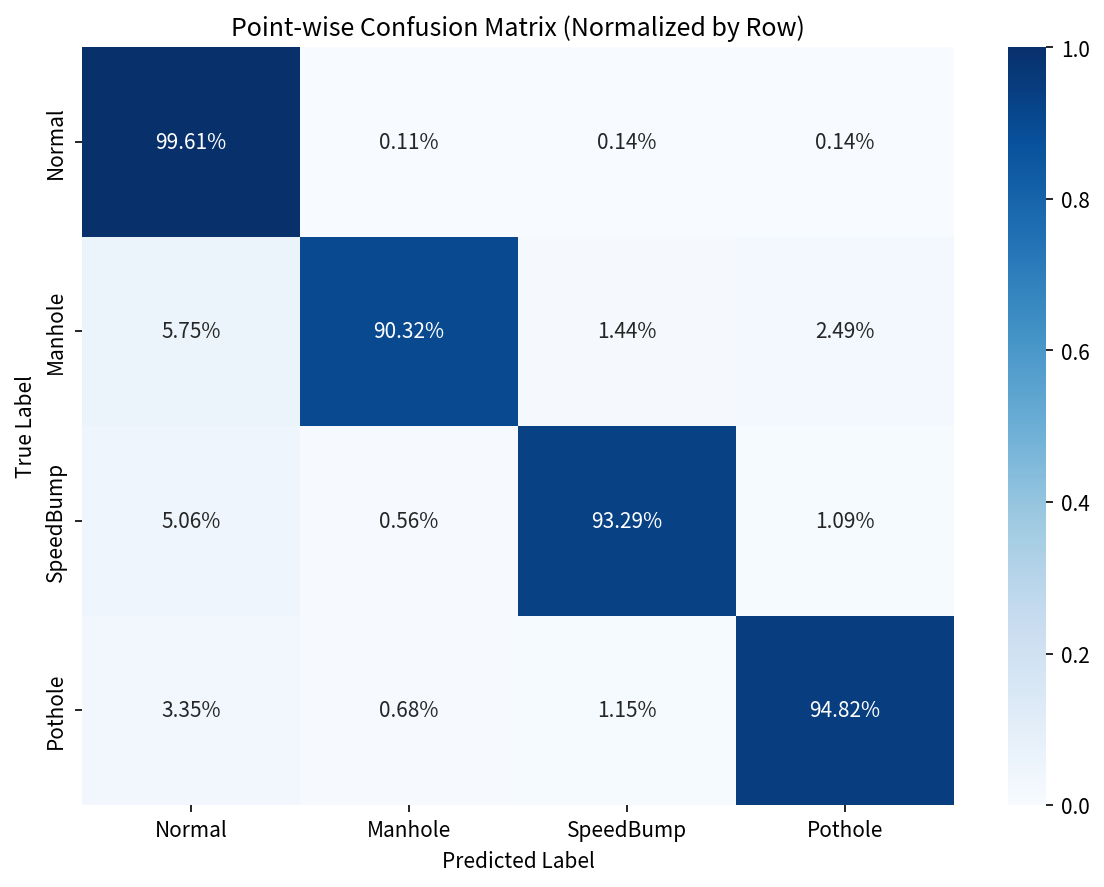}
	\label{subfig:fig20b}
	}
	%\quad%\hspace{10mm}
	\subfigure[CNN-Transformer + Point-wise.]{
	\includegraphics[width=0.22\linewidth]{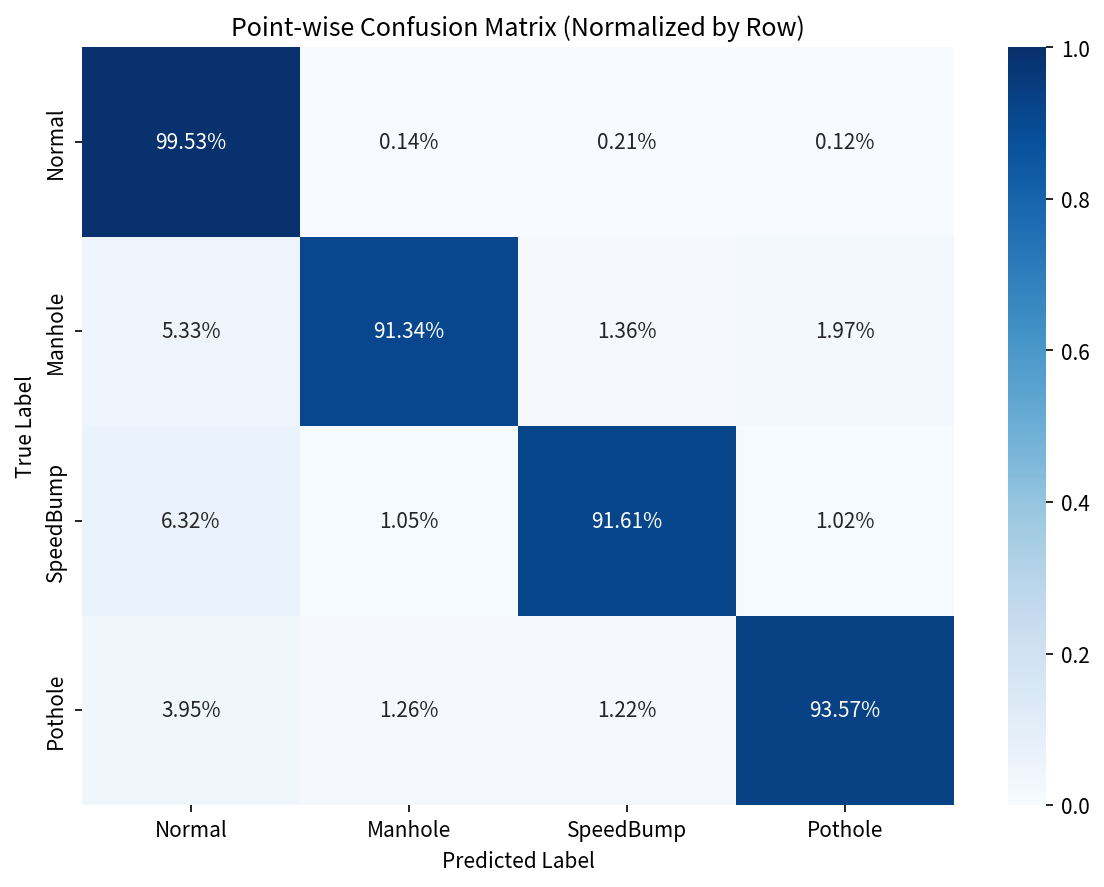}
	\label{subfig:fig20c}
	}
	%\quad%\hspace{10mm}
	\subfigure[Transformer + Point-wise.]{
	\includegraphics[width=0.22\linewidth]{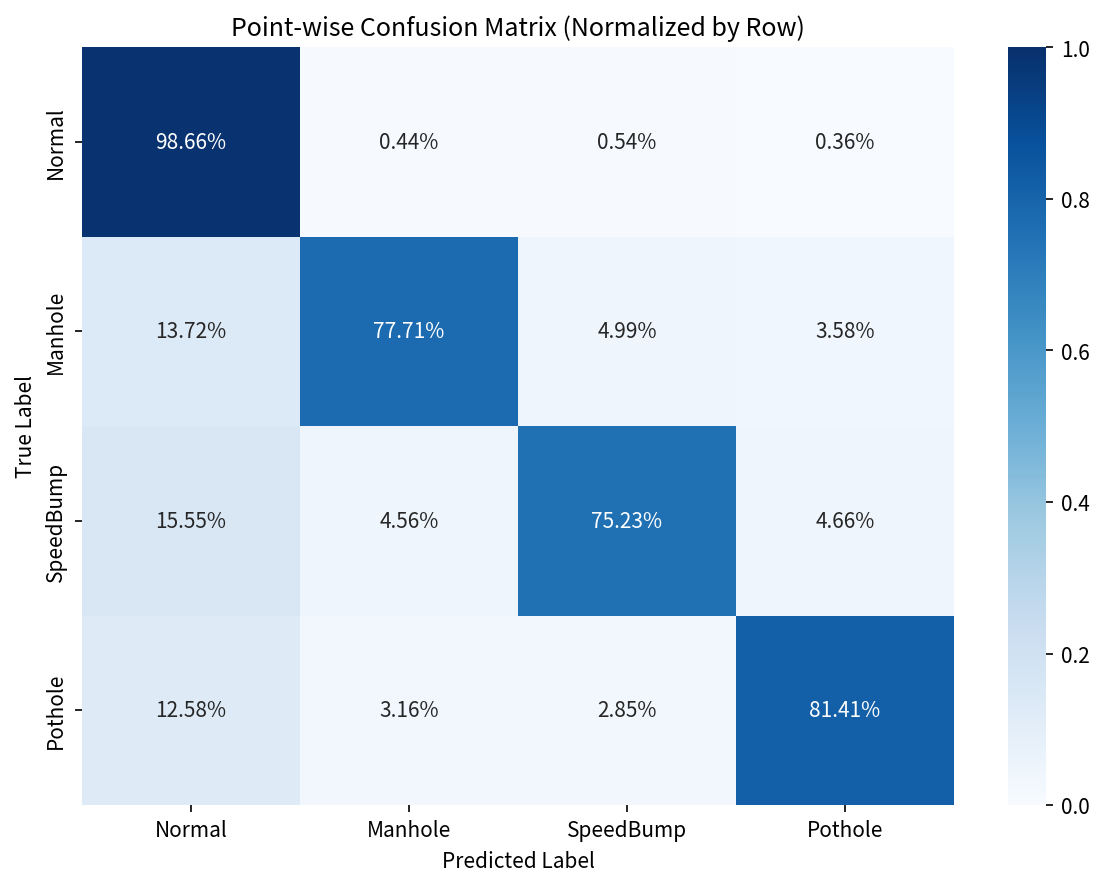}
	\label{subfig:fig20d}
	}	
	%\quad%\hspace{10mm}
	\subfigure[U-Net + Event-wise.]{
	\includegraphics[width=0.22\linewidth]{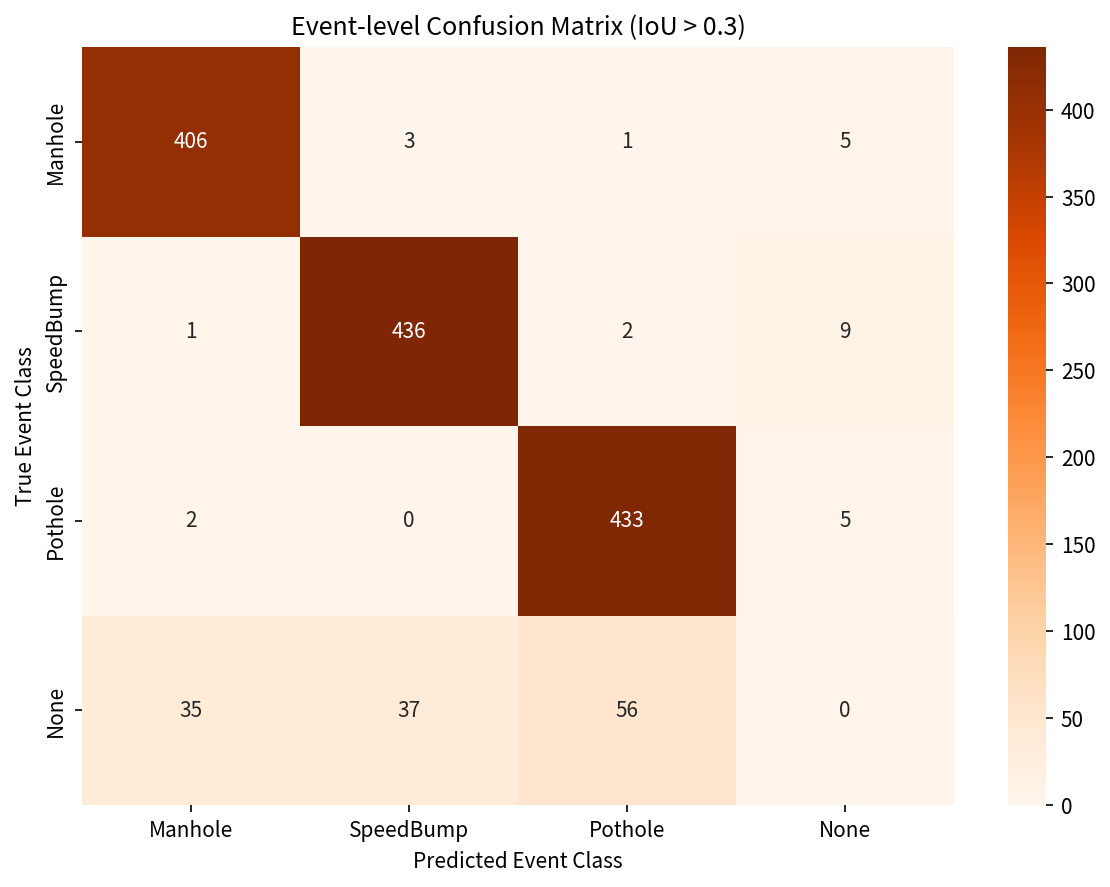}
	\label{subfig:fig21a}
	}
	%\quad%\hspace{10mm}
	\subfigure[CNN-LSTM + Event-wise.]{
	\includegraphics[width=0.22\linewidth]{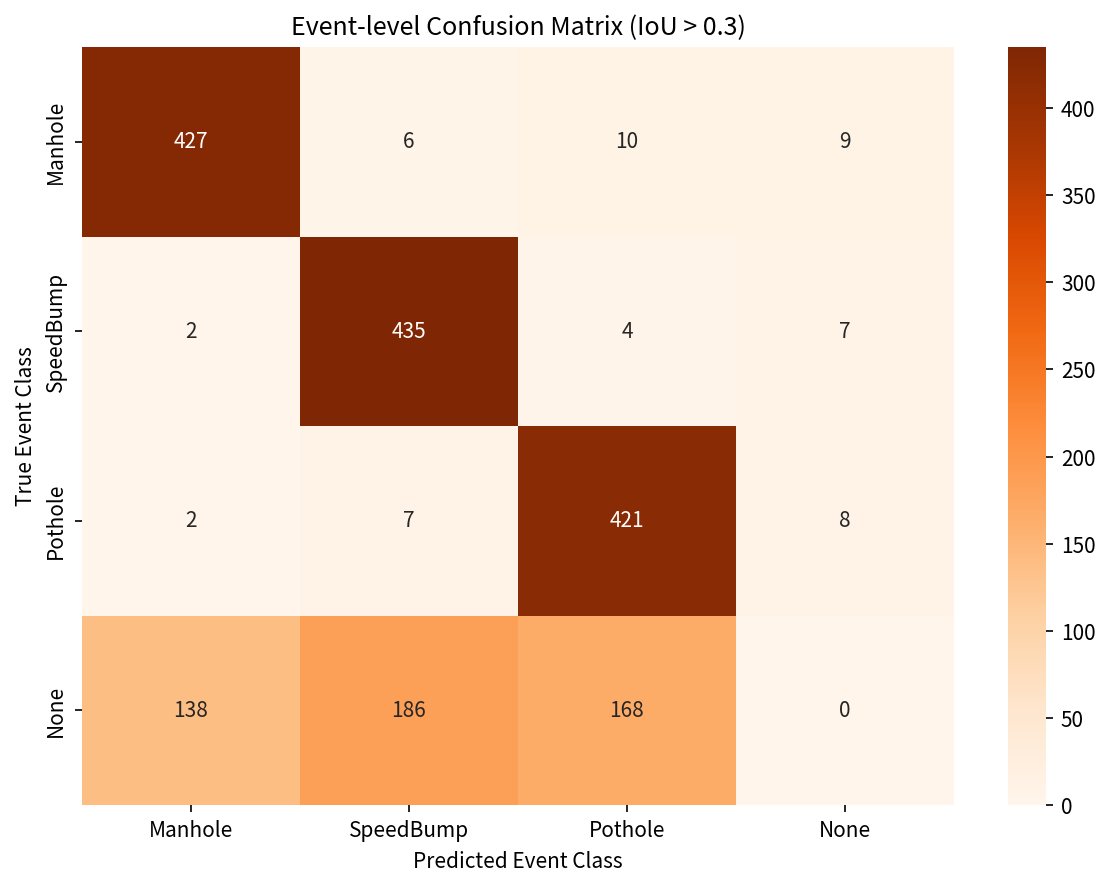}
	\label{subfig:fig21b}
	}
	%\quad%\hspace{10mm}
	\subfigure[CNN-Transformer + Event-wise.]{
	\includegraphics[width=0.22\linewidth]{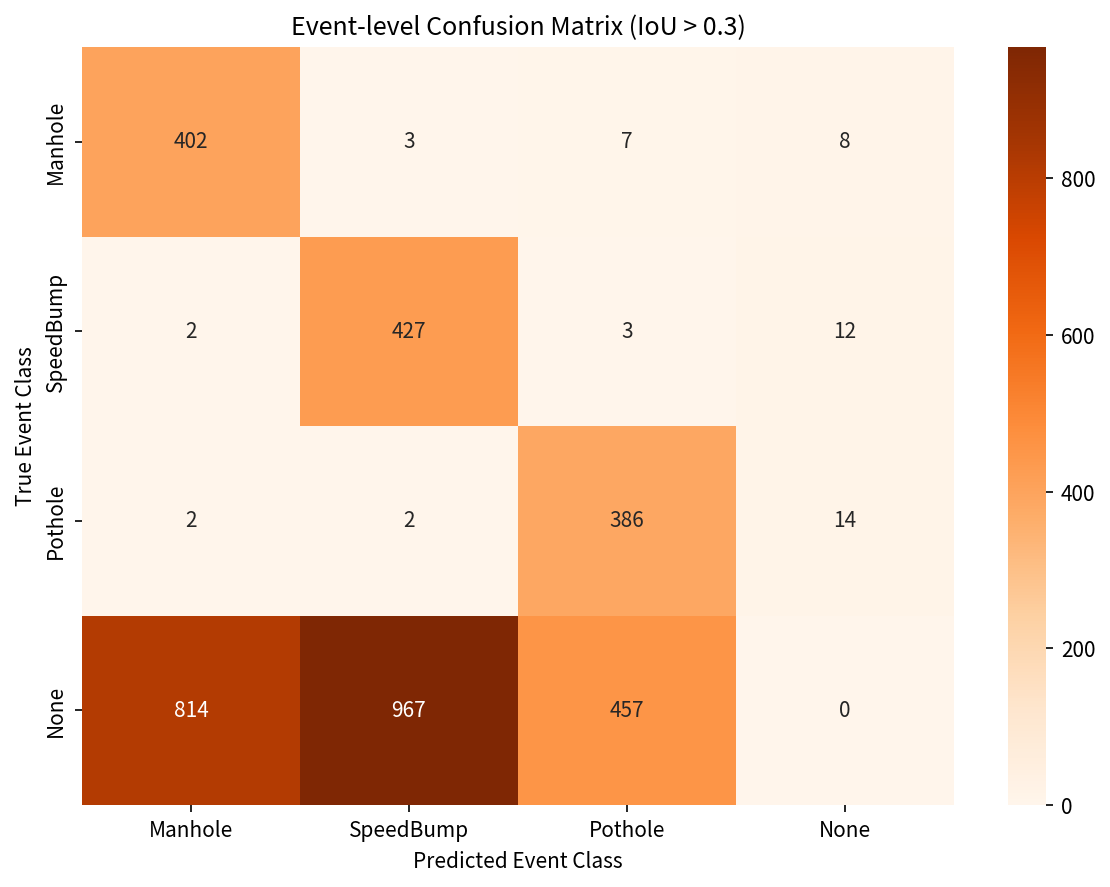}
	\label{subfig:fig21c}
	}
	%\quad%\hspace{10mm}
	\subfigure[Transformer + Event-wise.]{
	\includegraphics[width=0.22\linewidth]{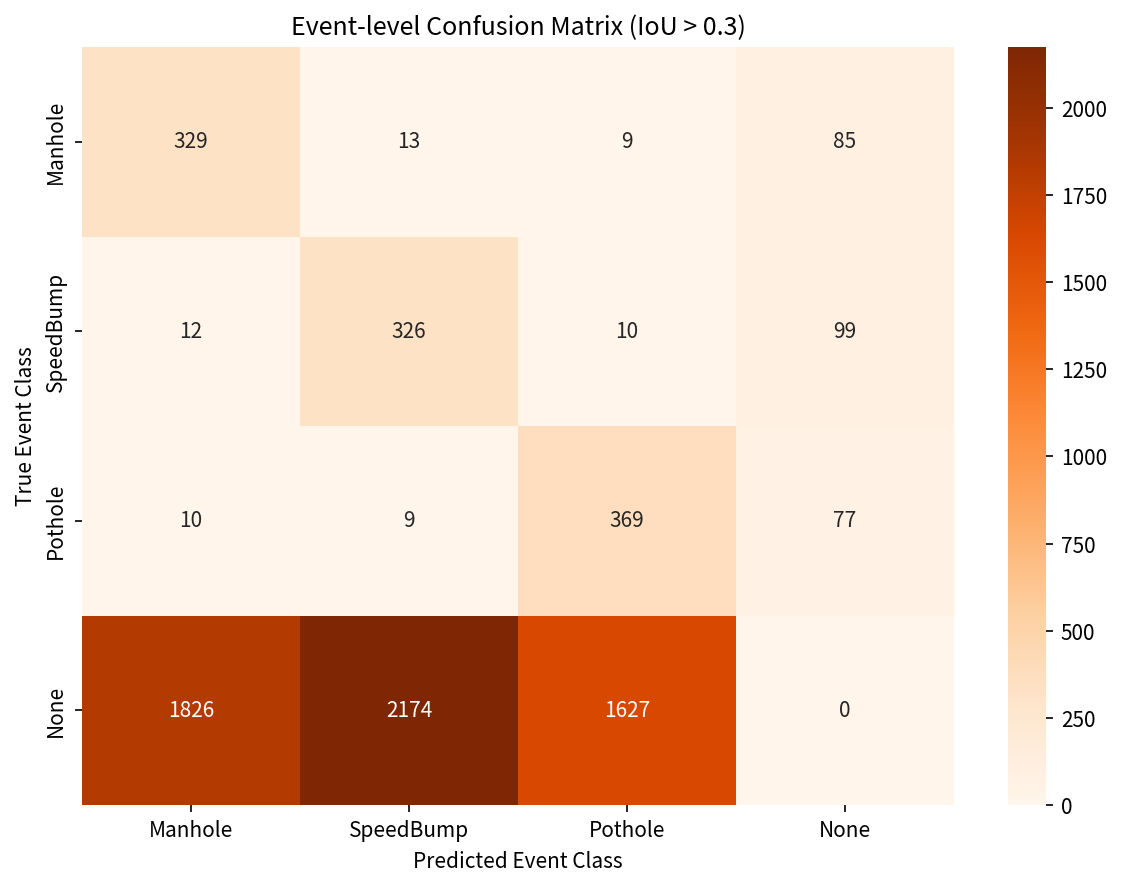}
	\label{subfig:fig21d}
	}
	\caption{Point-wise and event-wise confusion matrices under centralized training.}
	\label{fig:cm_centralized}
\end{figure*}
%-------------------

Comparing Tables~\ref{tab:centralized_core_metrics} and~\ref{tab:federated_core_metrics}, the event-level macro-average F1-score of U-Net decreases from 0.9426 under centralized training to 0.9244 under federated learning. This slight gap is acceptable because federated training does not directly centralize all local data and must handle heterogeneous vehicle distributions. The results validate the feasibility of federated temporal segmentation for multi-vehicle pothole detection.

%-------------------
\subsection{Point-Wise and Event-Wise Analysis}

The comparison between point-wise and event-wise metrics reveals an important characteristic of this task. Since normal background points dominate the vibration sequences, high point-wise accuracy may overestimate model performance. For example, under centralized training, Transformer achieves a point-wise accuracy of 0.9777 but only an event-level macro-average F1-score of 0.2584. Under federated learning, its event-level macro-average F1-score further drops to 0.2167. This indicates that the model correctly predicts many background points but fails to detect complete road events.

The proposed U-Net achieves a better balance between point-wise classification and event-wise localization. Its advantage comes from three aspects. First, the encoder extracts multi-scale temporal features from vibration sequences. Second, the decoder restores temporal resolution for dense prediction. Third, skip connections preserve local boundary information that may be lost during downsampling. These properties make U-Net more suitable for detecting short and localized vibration events.

The event-level IoU results further confirm this conclusion. The proposed U-Net not only predicts correct event categories but also aligns predicted intervals well with ground-truth intervals. This is particularly important for practical road maintenance, where inaccurate event boundaries may lead to unreliable pothole localization and reporting.

\begin{figure*}[!t]
	\centering 
	\subfigure[U-Net + Point-wise.]{
	\includegraphics[width=0.22\linewidth]{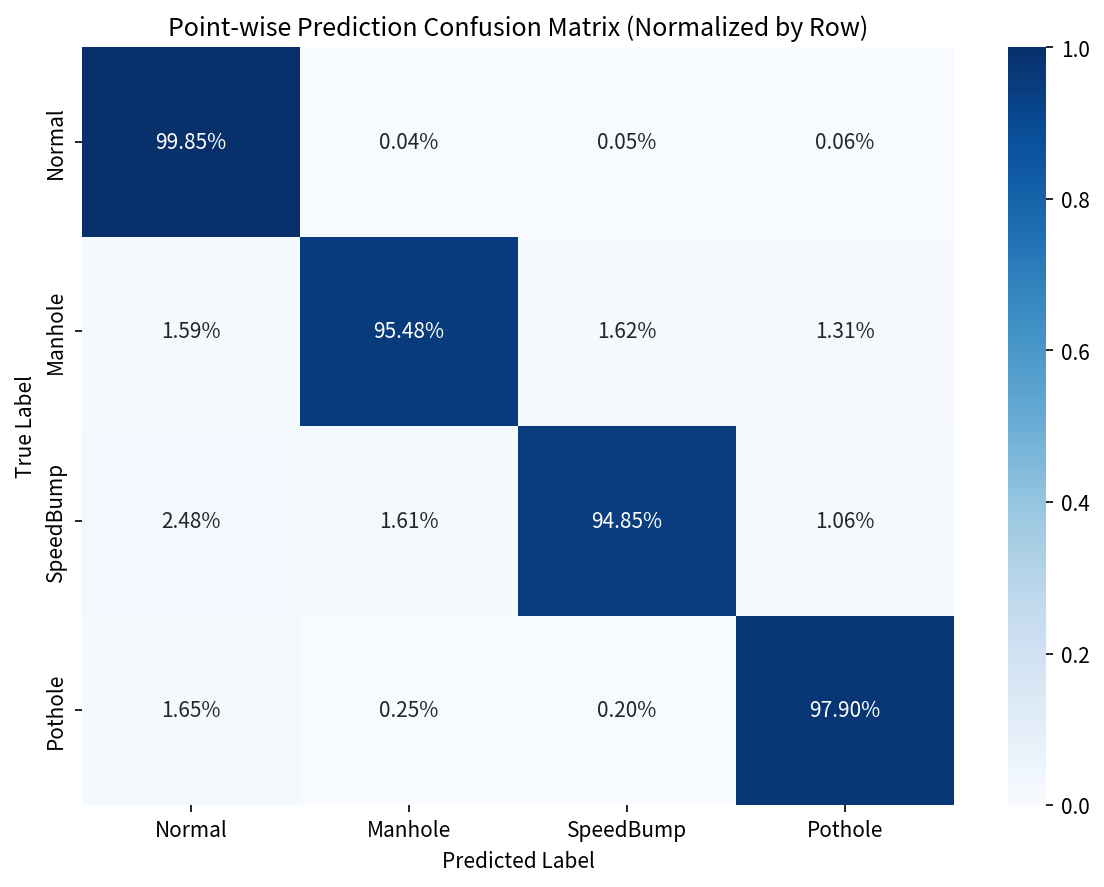}
	\label{subfig:fig22a}
	}
	%\quad%\hspace{10mm}
	\subfigure[CNN-LSTM + Point-wise.]{
	\includegraphics[width=0.22\linewidth]{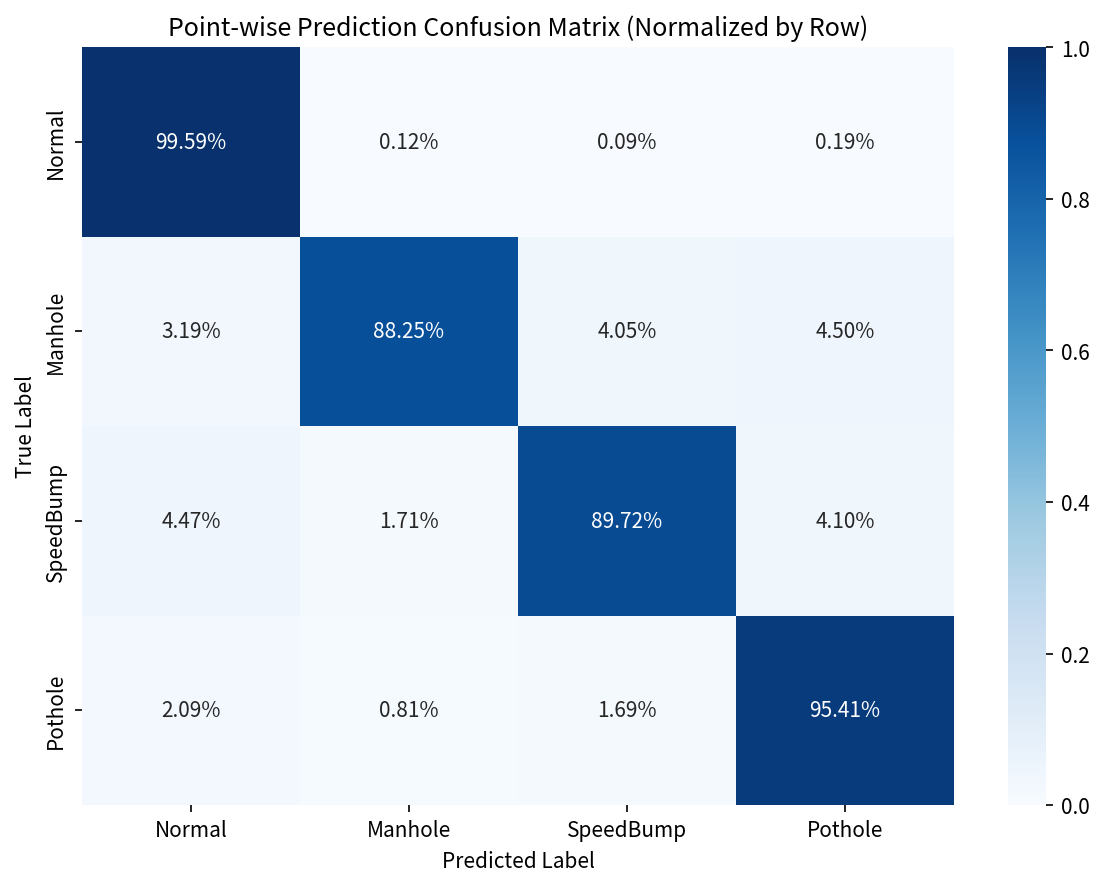}
	\label{subfig:fig22b}
	}
	%\quad%\hspace{10mm}
	\subfigure[CNN-Transformer + Point-wise.]{
	\includegraphics[width=0.22\linewidth]{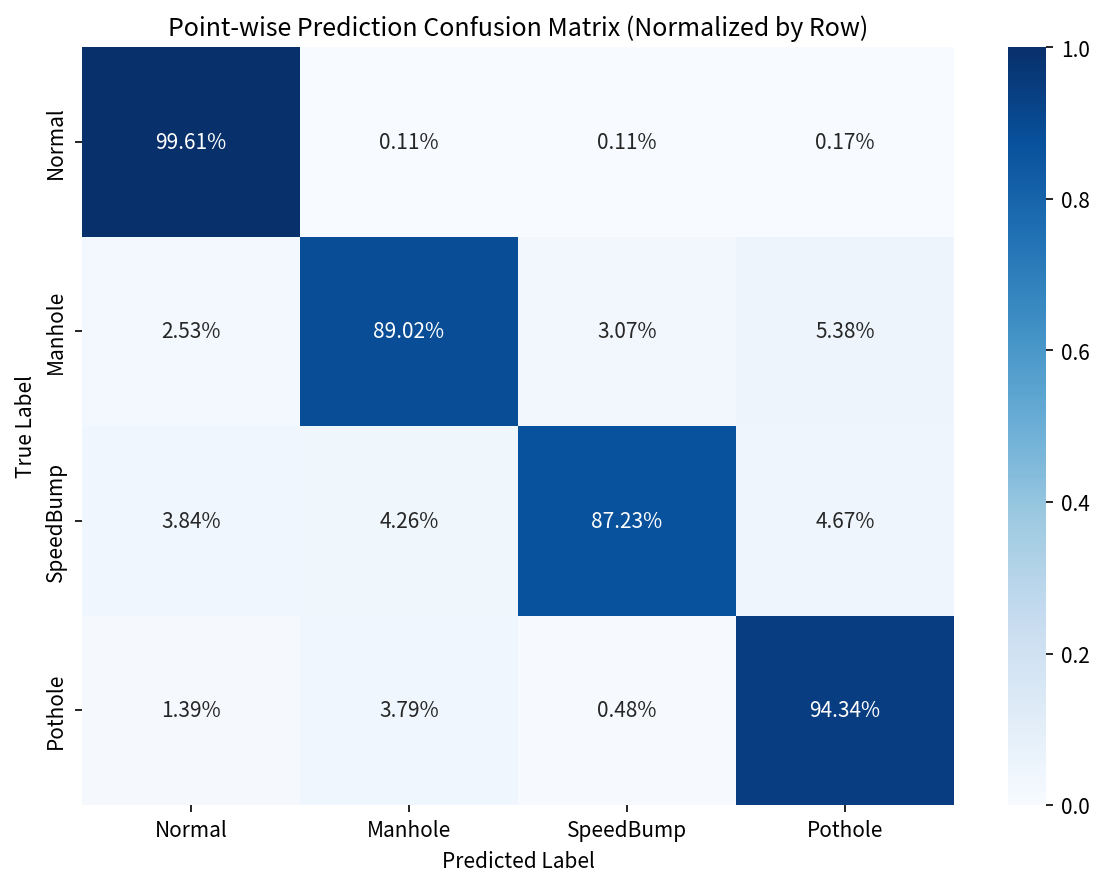}
	\label{subfig:fig22c}
	}
	%\quad%\hspace{10mm}
	\subfigure[Transformer + Point-wise.]{
	\includegraphics[width=0.22\linewidth]{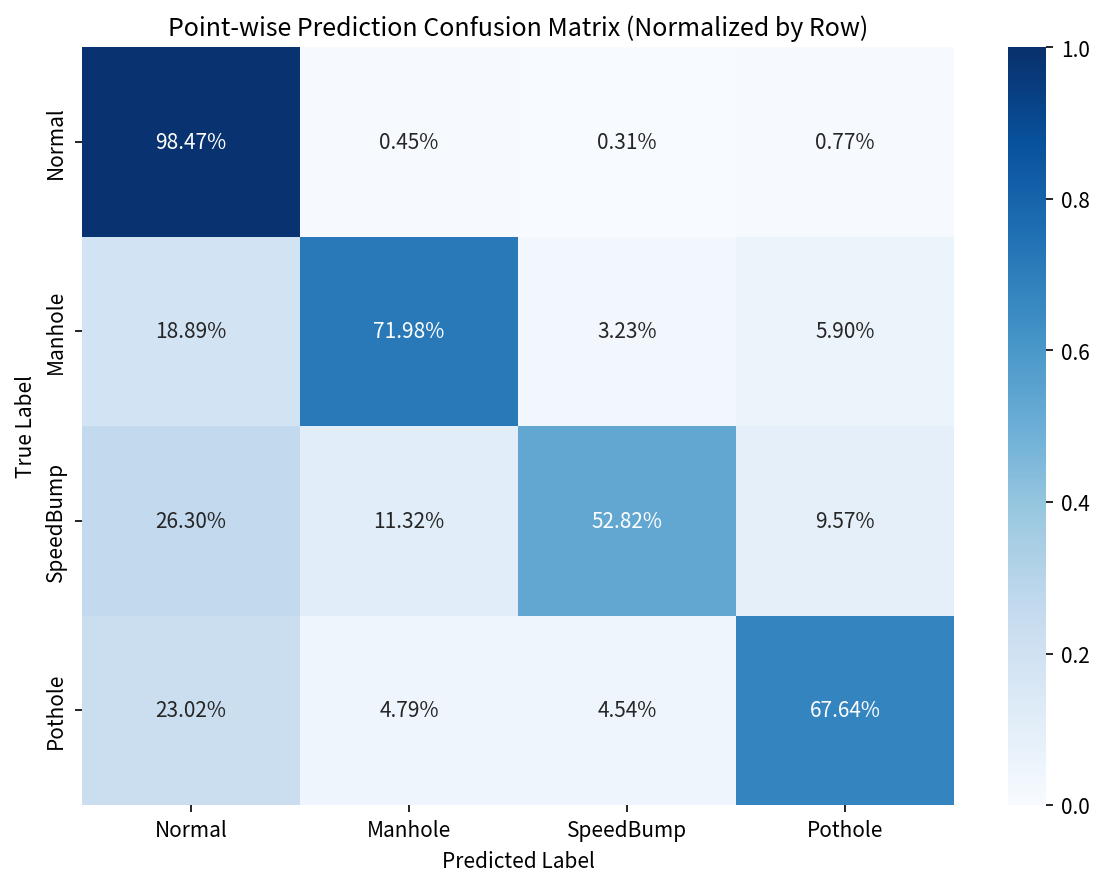}
	\label{subfig:fig22d}
	}
	%\quad%\hspace{10mm}
	\subfigure[U-Net + Point-wise.]{
	\includegraphics[width=0.22\linewidth]{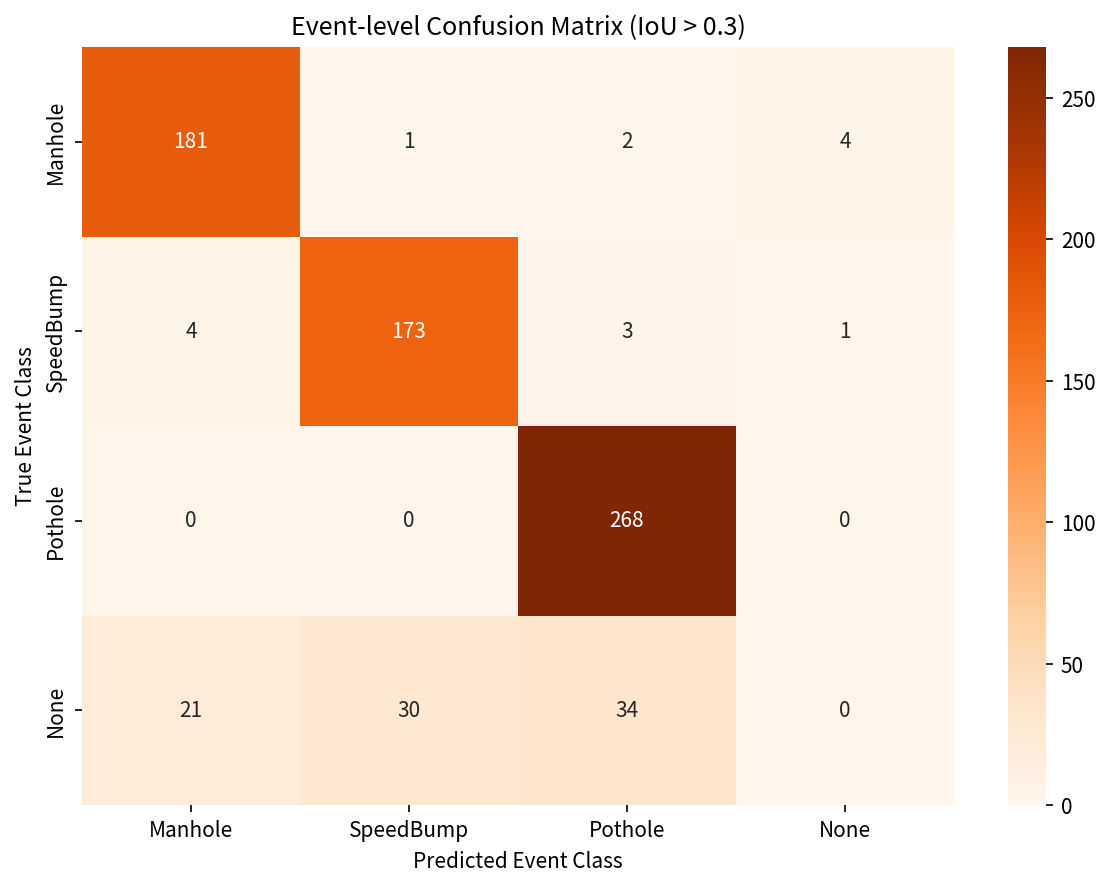}
	\label{subfig:fig23a}
	}
	%\quad%\hspace{10mm}
	\subfigure[CNN-LSTM + Point-wise.]{
	\includegraphics[width=0.22\linewidth]{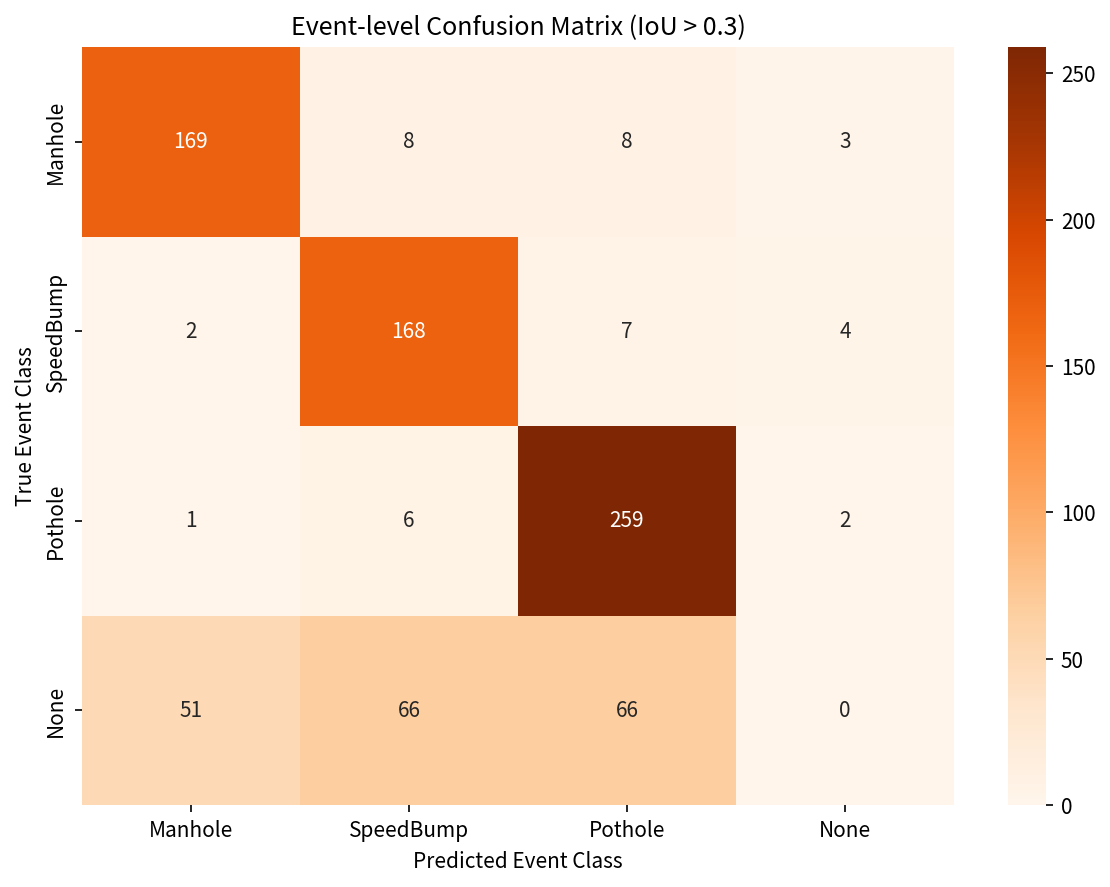}
	\label{subfig:fig23b}
	}
	%\quad%\hspace{10mm}
	\subfigure[CNN-Transformer + Point-wise.]{
	\includegraphics[width=0.22\linewidth]{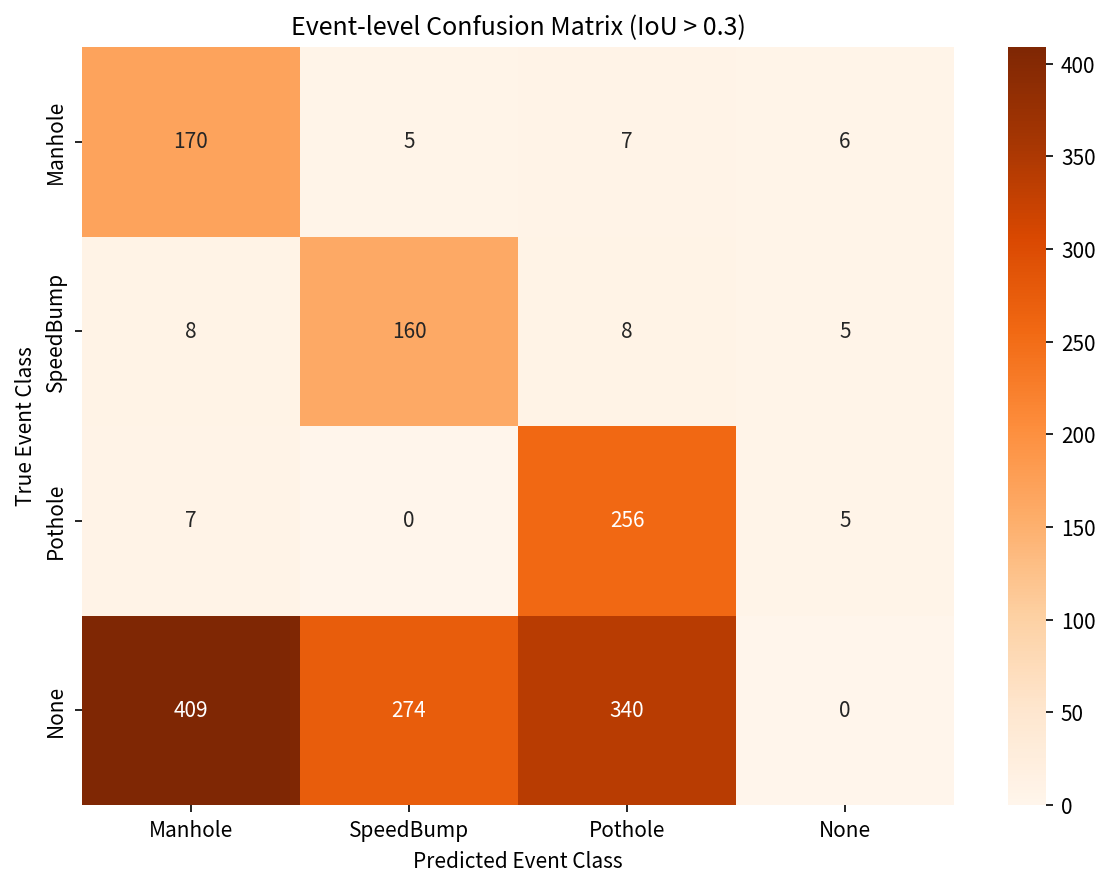}
	\label{subfig:fig23c}
	}
	%\quad%\hspace{10mm}
	\subfigure[Transformer + Point-wise.]{
	\includegraphics[width=0.22\linewidth]{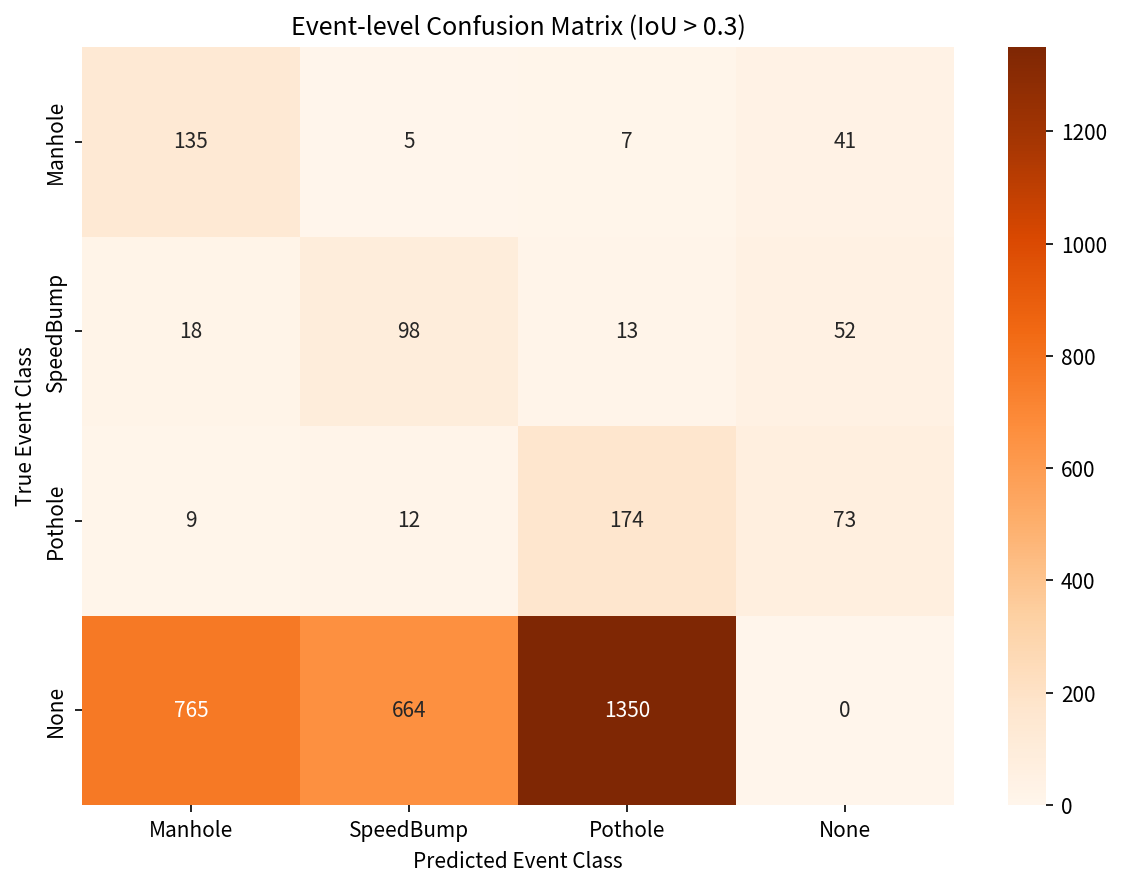}
	\label{subfig:fig23d}
	}	
	\caption{Point-wise and event-wise confusion matrices under federated learning.}
	\label{fig:cm_federated}
\end{figure*}
%-------------------

\subsection{Confusion Matrix and Visualization Analysis}

To further analyze category-level behavior, we examine point-wise and event-wise confusion matrices under centralized and federated settings. 
Fig.~\ref{fig:cm_centralized} and Fig.~\ref{fig:cm_federated}
show that the proposed U-Net produces clearer diagonal patterns than the baselines, especially for minority abnormal event classes.

The confusion matrices indicate that the main challenge is not distinguishing normal background from abnormal vibration, but distinguishing among potholes, manholes, and speed bumps. These events may produce similar acceleration peaks under certain vehicle speeds and road conditions. Compared with the baselines, U-Net reduces confusion among abnormal classes by combining local temporal details with multi-scale contextual features.

\subsection{Prediction Visualization and Post-Processing Analysis}

\begin{figure*}[t]
    \centering
    \includegraphics[width=0.8\linewidth]{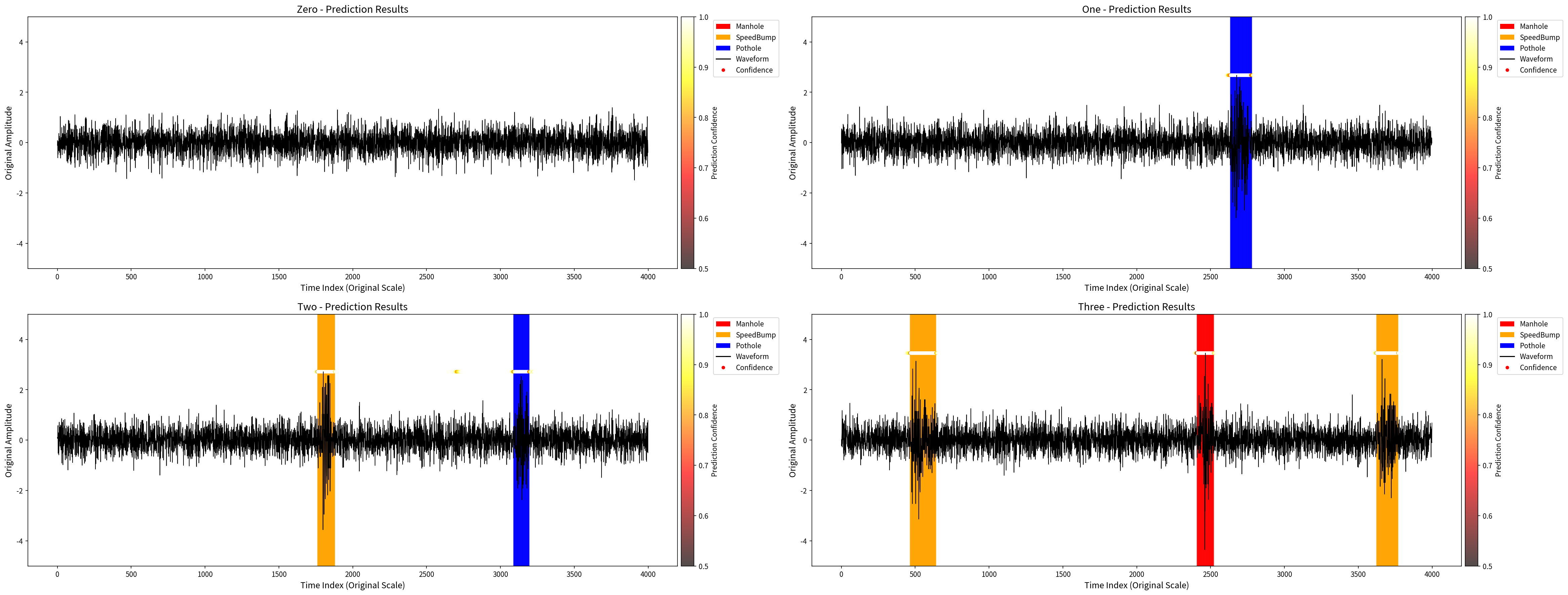}
    \caption{Visualization of prediction results. Abnormal regions are marked by color blocks, and point-wise classification confidence is displayed above the waveform.}
    \label{fig:visual_inf}
\end{figure*}

Fig.~\ref{fig:visual_inf} visualizes representative prediction results of the proposed 1D Attention U-Net. The waveform, predicted abnormal regions, and point-wise confidence are shown on the same time axis. The model can locate abnormal vibration intervals and assign corresponding road-event labels, showing its ability to perform both temporal segmentation and event classification.

\begin{figure}[t]
    \centering
    \includegraphics[width=0.7\linewidth]{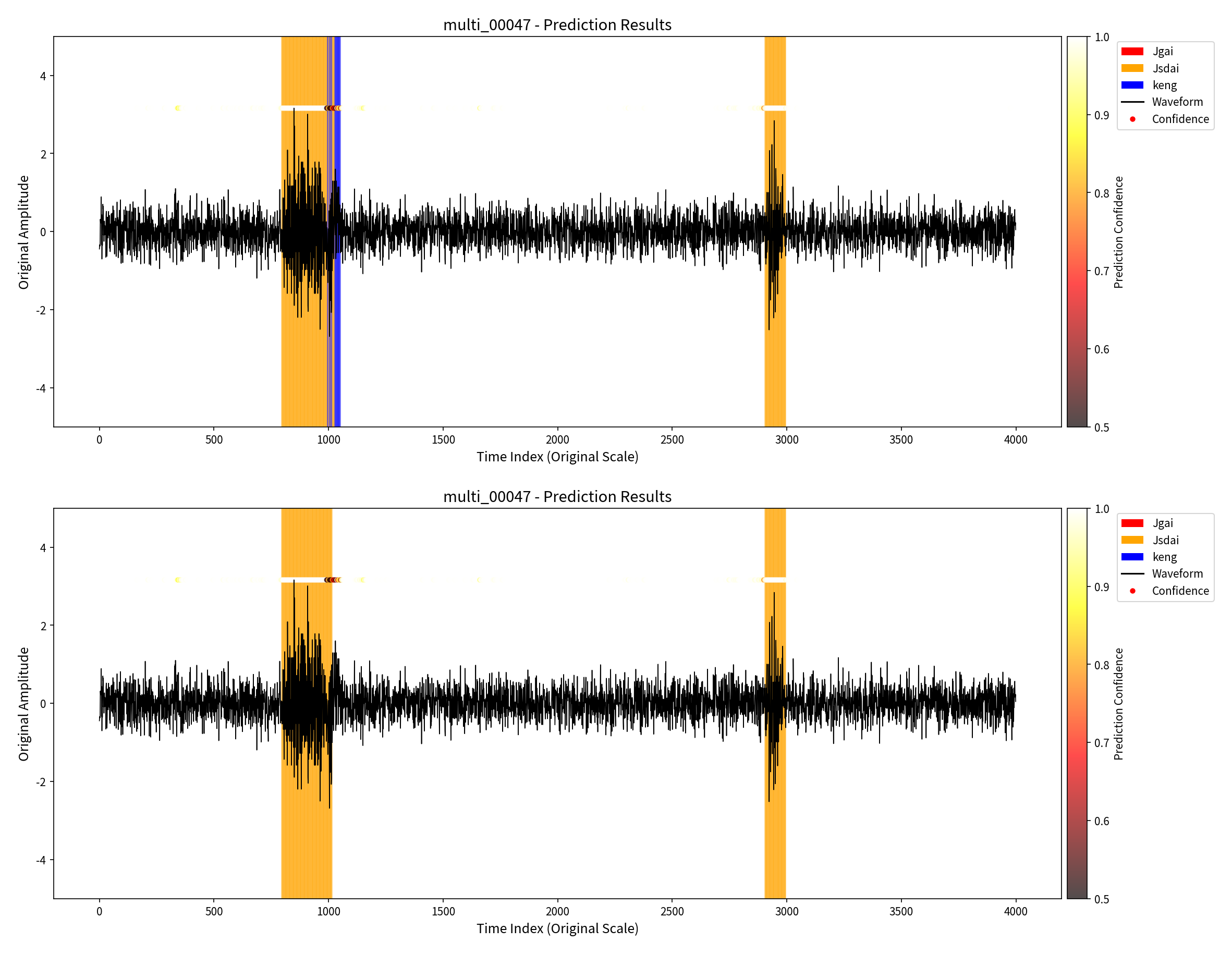}
    \caption{Visualization of post-processing effects. The upper graph shows the result before filtering, and the lower graph shows the result after filtering.}
    \label{fig:visual_filter}
\end{figure}

Fig.~\ref{fig:visual_filter} shows the effect of post-processing. The raw prediction contains several isolated or low-confidence misclassified fragments, while mode filtering and minimum length filtering remove these noisy segments and produce more continuous event regions. This improves the reliability of event-level pothole reports.

These visualization results indicate that the proposed model can accurately locate abnormal intervals, distinguish road-event categories, and generate more stable event-level outputs after post-processing.

\section{Conclusion and Future Work}

This paper proposed an edge-cloud collaborative framework for road pothole detection using vehicle-mounted vibration sensing. The framework decomposes pothole detection into onboard candidate event screening and server-side fine-grained temporal segmentation. At the vehicle side, a GMM-based background vibration model is used to adaptively detect abnormal vibration events from continuous acceleration streams, thereby reducing redundant transmission of smooth-road data. At the server side, a 1D Attention U-Net is developed to perform point-wise temporal segmentation over candidate event sequences, and the model is further trained under a federated learning framework to exploit distributed multi-vehicle data under non-IID conditions. Experimental results demonstrate that the proposed framework can effectively distinguish potholes from vibration-similar road events, improve event-level localization, and maintain robust performance under both centralized and federated settings.

Future work will further improve the system in several directions. First, we will validate the framework on larger-scale datasets collected from more cities, vehicle types, road materials, and weather conditions to evaluate its long-term generalization ability. 
Second, communication-efficient and personalized federated learning strategies will be explored to better handle unstable vehicular networks. 
Finally, GPS uncertainty, repeated reports from multiple vehicles, and spatial-temporal event fusion will be further studied to generate more accurate and reliable pothole maintenance reports for real-world deployment.
%-------------------

%-------------------
%\bibliographystyle{IEEEtran}
%\bibliography{references}

% Generated by IEEEtran.bst, version: 1.14 (2015/08/26)

%-------------------

\end{document}